\documentclass[superscriptaddress,aps,prx,twocolumn]{revtex4-2}

\usepackage{url}
\usepackage{amssymb}% provides \mathbb (used for the identity matrix)
\usepackage{physics}
\usepackage{graphicx}% Include figure files
\usepackage{dcolumn}% Align table columns on decimal point
\usepackage{bm}% bold math
\usepackage{hyperref}% add hypertext capabilities
\usepackage{siunitx}
\usepackage{booktabs}
\usepackage{wrapfig}
\usepackage{lipsum}
\usepackage[none]{hyphenat}
\begin{document}

\preprint{APS/123-QED}

\title{Gate Control of g-factor in Germanium Quantum Dots: A Strain-Based Explanation}% Force line breaks with \\

\author{Mu Niu}
\affiliation{Department of Physics and Astronomy, University of California, Los Angeles, 90095, California, USA}

\author{Adrian Culver}
\affiliation{Center for Quantum Science and Engineering, University of California, Los Angeles, 90095, California, USA}
\affiliation{Department of Electrical and Computer Engineering, University of California, Los Angeles, 90095, California, USA}

\author{Johnathan Bryan}
\affiliation{Department of Physics and Astronomy, University of California, Los Angeles, 90095, California, USA}

\author{Chris Anderson}
\affiliation{Center for Quantum Science and Engineering, University of California, Los Angeles, 90095, California, USA}
\affiliation{Department of Mathematics, University of California, Los Angeles, 90095, California, USA}

\author{Mark Gyure}
\affiliation{Center for Quantum Science and Engineering, University of California, Los Angeles, 90095, California, USA}
\affiliation{Department of Electrical and Computer Engineering, University of California, Los Angeles, 90095, California, USA}

\author{HongWen Jiang}
\affiliation{Department of Physics and Astronomy, University of California, Los Angeles, 90095, California, USA}
\affiliation{Center for Quantum Science and Engineering, University of California, Los Angeles, 90095, California, USA}


\begin{abstract}
The g-factor is a key parameter governing the behavior of semiconductor spin qubits, as it directly determines the qubit frequency and its sensitivity to electrical and magnetic noise. Recent experiments in germanium quantum dots have revealed large g-factor variations under small gate voltage changes, indicating a strong coupling between electrostatics and spin properties. Here, we present a quantitative explanation based on strain-induced g-tensor modulation. By combining finite-element simulations of inhomogeneous strain with quantum calculations of hole wavefunctions, we show that device-induced strain produces spatially varying g-tensors. Gate voltages shift the quantum dot within this landscape, leading to substantial changes in the effective g-factor. Our results may account for the experimentally observed tunability and highlight the importance of in-plane g-tensor variations. This work establishes a direct link between strain, electrostatic control, and qubit performance in germanium spin qubits.
\end{abstract}
\maketitle

%\tableofcontents

\section{Introduction}

%Germanium quantum dots are promising candidates for quantum information processing due to their advantageous properties. Germanium hole spin states exhibit long coherence times and fast rotation speeds, both of which are essential for stable qubit performance \cite{watzinger2018germanium, jirovec2021singlet}. A key advantage of germanium is that its two topmost valence bands—the Heavy Hole (HH) and Light Hole (LH) bands—are well separated from other bands. These bands also have a light effective mass and lack valley degeneracy, making them ideal for spin encoding \cite{chekhovich2015suppression}. Additionally, the strong spin-orbit interaction (SOI) in germanium hole spins allows for all-electric qubit control, eliminating the need for micromagnets \cite{Scappucci, Kloeffel}.%

Germanium hole spin qubits have emerged as a leading platform for semiconductor quantum information processing, owing to their strong spin–orbit interaction, low effective mass, and absence of valley degeneracy. These properties enable fast, all-electric qubit control without the need for micromagnets, while maintaining long coherence times~\cite{watzinger2018germanium, chekhovich2015suppression, Scappucci, Kloeffel,zajac2018resonantly, petta2005coherent, Burkard2023SpinQubits}. In particular, the anisotropic and site-dependent g-tensor in Ge quantum dots plays a central role in determining both the qubit frequency and its susceptibility to noise, making it a key parameter for device performance and scalability~\cite{jirovec2021singlet}.

Recent experiments have revealed that the g-factor in germanium quantum dots can exhibit remarkably strong dependence on gate voltages. Notably, Rooney \textit{et al.} demonstrated that the singlet–triplet qubit frequency can change by nearly an order of magnitude with only a ~12 mV variation in a barrier gate voltage~\cite{rooney2023gate}. Such sensitivity suggests that small electrostatic changes can significantly modify the local electronic environment of the quantum dots, leading to large variations in the effective Zeeman splitting. While this tunability offers new opportunities for qubit control~\cite{hendrickx2024sweet, michal2021longitudinal}, it also raises fundamental questions about the microscopic origin of these g-factor variations.

%Thermal contraction-induced strain plays a major role in the inhomogeneity of the g-tensor. Quantum dot devices typically consist of materials with different thermal expansion coefficients, such as aluminum or gold gates and quantum dot layers made of Ge, Si, or GaAs \cite{mauro2024strain}. As these devices are cooled from room to cryogenic temperatures, the differing thermal contraction rates generate inhomogeneous strain. This strain alters the g-tensor locally, leading to spatially varying corrections \cite{abadillo-uriel2023hole}.

A compelling theoretical framework for understanding such behavior has been developed by Abadillo-Uriel \textit{et al.}, who showed that inhomogeneous strain fields can strongly influence both spin–orbit interactions and the g-tensor in hole-based systems. In their work, strain gradients—arising naturally from fabrication processes and thermal contraction—induce corrections to the g-tensor and generate additional spin–orbit coupling mechanisms. Crucially, because the g-tensor depends on the spatial distribution of strain, any displacement of the quantum dot (for example, due to gate voltage changes) can lead to substantial modulation of the effective g-factor~\cite{abadillo-uriel2023hole}. This establishes a direct link between electrostatic control and spin properties via the underlying strain landscape.

%Despite extensive research on the anisotropic g-tensor in semiconductors, several critical gaps remain. One major gap is the lack of a quantitative description of how strain-induced corrections to the g-tensor are distributed throughout a quantum dot device. Additionally, the effect of gate voltage on the electron density distribution—and how this, in turn, modifies the g-tensor experienced by the quantum dot—has received limited investigation. Importantly, the interplay between strain, gate voltage, and g-tensor corrections has not been systematically analyzed, leaving a significant gap in understanding their combined impact on qubit performance. Further, Existing studies on the anisotropic g-tensor emphasize differences between in-plane and out-of-plane components. Less attention is given to the variation within in-plane components, although these variations may appear minor to out-of-plane components, their effects could become significant when the device operates under in-plane magnetic field. Such subtle changes are key to explaining the experimentally observed modulation of qubit frequency and dephasing time as a function of gate voltage, and addressing these finer details is essential for developing more precise and effective qubit control strategies. 

Despite this progress, a quantitative and device-relevant understanding of how gate-induced dot motion couples to spatially inhomogeneous g-tensor fields remains incomplete. In particular, it is not yet fully understood how realistic strain distributions, combined with the finite extent of the hole wavefunction, give rise to the large g-factor variations observed experimentally. Furthermore, the role of in-plane g-tensor components—often neglected due to their smaller magnitude—may become significant when the magnetic field orientation and device geometry are taken into account.

%This study aims to provide a systematic and quantitative analysis of strain-induced g-tensor corrections caused by thermal contraction and gate modulation of g-factor. First, we simulate the strain distribution by solving the thermo-elasticity equation using FEniCS, a Finite Element Method (FEM) package for Python \cite{Alnaes2015FEniCS, Logg2012Automated, Logg2010DOLFIN}. The resulting strain map is used to calculate the spatial distribution of strain-induced corrections to the g-tensor. Next, we compute the hole density distribution at various gate voltages using MaSQE. Finally, we combine these results to determine the effective g-factor for a single dot and for a singlet-triplet qubit under different in-plane magnetic field orientations.

In this work, we address these questions by developing a comprehensive framework that connects strain-induced g-tensor inhomogeneity, gate-controlled charge distribution, and the resulting effective g-factor of quantum dot qubits. Building on the theoretical foundation of strain-induced spin–orbit coupling and g-tensor modulation, we incorporate realistic simulations of strain fields and hole wavefunctions to evaluate how the effective g-factor evolves under gate voltage tuning. We show that the interplay between spatially varying g-tensor corrections and gate-induced dot displacement naturally explains the large tunability observed in experiments.

Our results provide a unified interpretation of g-factor modulation in germanium quantum dots and highlight the critical role of strain engineering in spin qubit devices. By elucidating the microscopic mechanisms behind g-tensor variability, this work offers new insights into optimizing qubit performance and designing more robust, controllable semiconductor spin qubits.

\section{Device Structure} \label{Device Structure}

\begin{figure}
    \centering
    \includegraphics[width=1\linewidth]{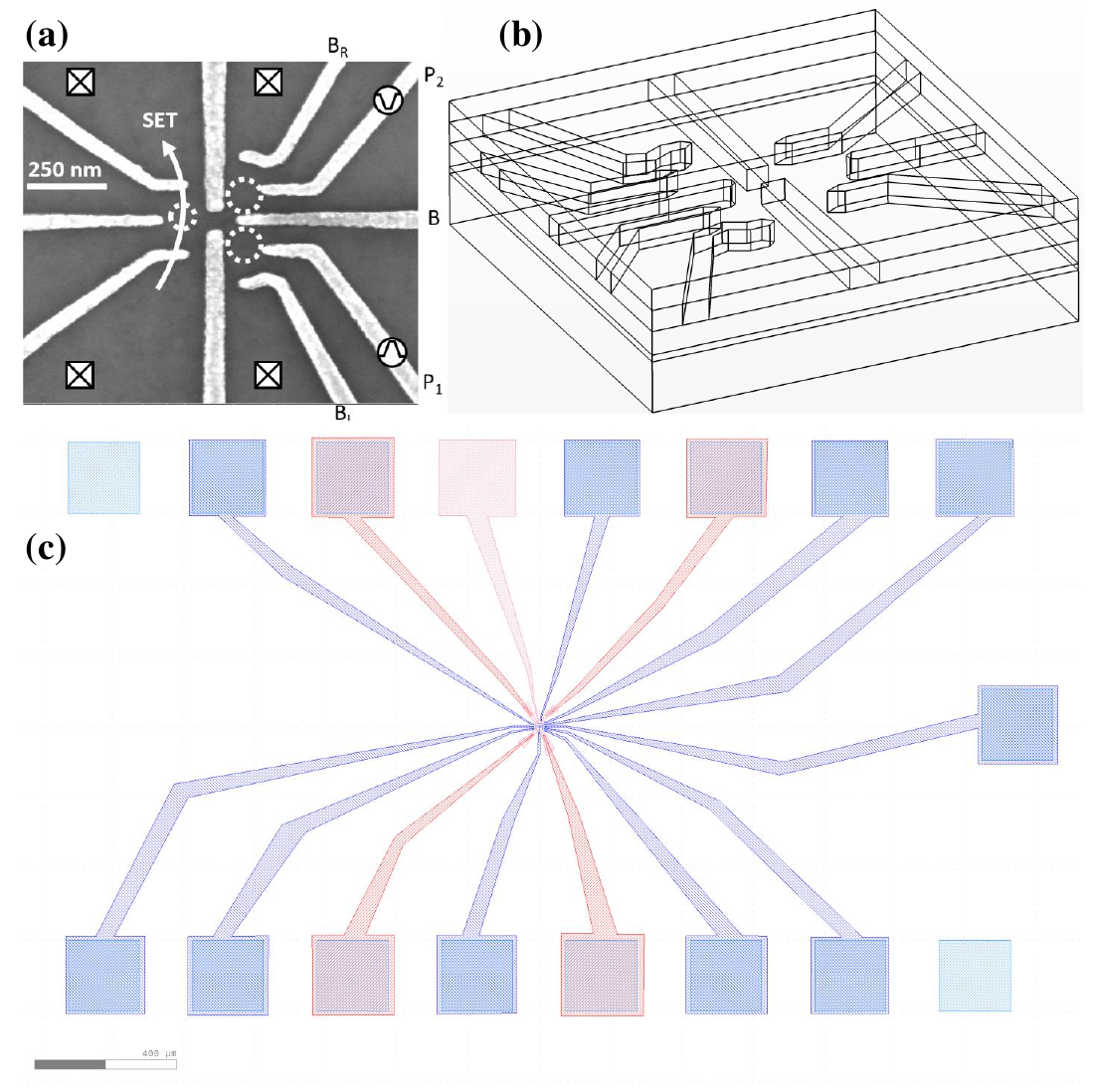}
    \caption{(a) SEM image of lithographically defined gates identical to the geometry used in this study. \(B_L\) and \(B_R\) are confinement gates, \(P_1\), \(P_2\), and \(B\), as a shorthand for Brm, are control gates, SET refers to single electron transistor which is responsible for readout. (b) View of the simulation geometry in Gmsh, detailing the layer structure of the device. From bottom to top: a \(120\ \text{nm}\) thick \(\text{Ge}_{0.8}\text{Si}_{0.2}\) layer, a \(16\ \text{nm}\) thick Ge layer, a \(55\ \text{nm}\) thick \(\text{Ge}_{0.8}\text{Si}_{0.2}\) layer, a \(50\ \text{nm}\) thick Al gate layer, and a \(50\ \text{nm}\) thick \(\text{Al}_2\text{O}_3\) cap layer. Areas not occupied by gates within the gate layer are also filled with \(\text{Al}_2\text{O}_3\). (c) View of the entire device in KLayout, which has a length scale of approximately \(10^2\ \mu\text{m}\), while the region selected for simulation has a length scale of around \(10^2\ \text{nm}\).}
    \label{Device}
\end{figure}

A typical quantum dot qubit device consists of a semiconductor substrate, an active quantum dot layer, and a set of overlying electrical gates. In our laboratory, the device is constructed as follows (from bottom to top): a \(120\,\text{nm}\)-thick \(\text{Ge}_{0.8}\text{Si}_{0.2}\) layer serves as the substrate, followed by a \(16\,\text{nm}\)-thick Ge layer containing the quantum dots. Above this, a \(55\,\text{nm}\)-thick \(\text{Ge}_{0.8}\text{Si}_{0.2}\) layer is deposited, then a \(50\,\text{nm}\)-thick Al gate layer, and finally a \(50\,\text{nm}\)-thick \(\text{Al}_2\text{O}_3\) cap. Regions within the gate layer not occupied by aluminum gates are also filled with \(\text{Al}_2\text{O}_3\). The gate configuration includes two barrier gates (\(B_L\) and \(B_R\)) that confine the quantum dot, and a central Brm gate that regulates coupling between two adjacent dots (see Fig.~\ref{Device} (a)).

For our FEM simulation, we aimed to accurately model the strain distribution in the vicinity of the quantum dots. Consequently, the simulation mesh geometry (see Fig.~\ref{Device} (b)) replicates our experimental device (Fig.~\ref{Device} (c)) only in the region near the dots. To minimize boundary effects in simulation, we included a sufficiently large volume in the simulation geometry surrounding the quantum dot region so that the dot remains well separated from the boundaries.

\section{Theory}

\subsection{Thermal Contraction Strain}

This section explains how strain arises from differential thermal contraction. We present the equations used for the simulation at the end. Consider a volume \(\Omega\); the total work done by the strain inside this volume is given by
\begin{eqnarray}
    \label{stress work}
    W = \int_\Omega \vb{\sigma(u)} : \vb{\delta \epsilon(u)} \, dV
\end{eqnarray}
where \(\sigma\) is the stress tensor, \(\epsilon\) is the strain tensor, and \(\delta \epsilon\) refers to the infinitesimal strain tensor. The symbol \(:\) denotes the double contraction between the stress and strain tensors, which effectively sums over the products of corresponding components of \(\sigma\) and \(\epsilon\). Suppose we change the temperature from the unstrained value $T_0$ to $T$, the thermal contraction stress tensor can be expressed in terms of the strain tensor as
\begin{eqnarray}
    \label{stress}
    \mathbf{\sigma} = \lambda \Tr(\mathbf{\epsilon}) \mathbf{I} + 2\mu \mathbf{\epsilon} - \alpha (3\lambda + 2\mu)(T-T_0)\mathbf{I}
\end{eqnarray}
where \(\lambda\) and \(\mu\) are the Lamé parameters and $\mathbf{I}$ is the identity.

In equilibrium, the total work must be balanced by the work done by external forces acting on the boundary of the domain \(\Omega\). Thus, if we perturb the displacement \(\vb{u}\) by an infinitesimal displacement \(\vb{v}\), we obtain the equation we use in this study:
\begin{eqnarray}
    \label{Elasticity}
    \forall \vb{v} \in V: \quad \int_\Omega \vb{\sigma(u)} : \vb{\epsilon(v)} \, d\Omega = \int_{\partial \Omega} \vb{T} \vdot \vb{v} \, dS
\end{eqnarray}
where \(\vb{T}\) represents the external force per unit area acting on the boundary \(\partial \Omega\). In this study, we take \(\Omega\) to be the volume of the entire geometry used in this simulation and assume no external force on its surface, i.e. \(\vb{T} = 0\). For detailed derivation, see Appendix \ref{Thermal Elasticity} \cite{landau1986theory}.

\subsection{Strain-Induced g-Tensor Correction}

The motion of a germanium hole spin under a uniform magnetic field \(\mathbf{B}\) is described by the Luttinger-Kohn Hamiltonian:
\begin{equation}
    H = H_k + H_\epsilon + H_z + V(\mathbf{r}) \mathbb{I},
\end{equation}
where \(H_k\) is the kinetic Hamiltonian, \(H_\epsilon\) represents the strain Hamiltonian, and \(H_z\) is the Zeeman Hamiltonian. The HH and LH states correspond to the z-component of the hole angular momentum \(J_z = \pm \frac{3}{2}\) and \(J_z = \pm \frac{1}{2}\), respectively \cite{Scappucci}. 

Using the angular momentum $J_z$ eigenstates \( \big\{ \ket{ + \frac{2}{3}} \ket{ + \frac{1}{3}} \ket{ - \frac{1}{3}} \ket{ - \frac{2}{3}} \big\} \) as the basis, the kinetic and strain Hamiltonians take the same form:
\begin{equation}
    H_{k/\epsilon} = \mqty(
    P+Q & -S & R & 0 \\ 
    -S^\dag & P-Q & 0 & R \\
    R^\dag & 0 & P-Q & S \\
    0 & R^\dag & S^\dag & P+Q).
\end{equation}
For the kinetic Hamiltonian, the \(P\), \(Q\), \(R\), and \(S\) variables are given by:
\begin{equation}
    \begin{aligned}
        P_k &= \frac{1}{2m_0} \gamma_1 (p_x^2 + p_y^2 + p_z^2), \\
        Q_k &= \frac{1}{2m_0} \gamma_2 (p_x^2 + p_y^2 - 2p_z^2), \\
        R_k &= \frac{1}{2m_0} \sqrt{3} \left[ -\gamma_2 (p_x^2 - p_y^2) + 2i \gamma_3 \{p_x, p_y\} \right], \\
        S_k &= \frac{1}{2m_0} 2 \sqrt{3} \gamma_3 \{p_x - i p_y, p_z\},
    \end{aligned}
\end{equation}
where \(\gamma_1\), \(\gamma_2\), and \(\gamma_3\) are the Luttinger parameters, determined entirely by the material properties. The terms \(p_x\), \(p_y\), and \(p_z\) represent the components of momentum, and \(\{ A,B \} = AB + BA\) denotes the anti-commutator. 

For the strain Hamiltonian, the \(P\), \(Q\), \(R\), and \(S\) variables are:
\begin{equation}
    \begin{aligned}
        P_{\epsilon} &= -a_v (\epsilon_{xx} + \epsilon_{yy} + \epsilon_{zz}), \\
        Q_{\epsilon} &= -\frac{1}{2} b_v (\epsilon_{xx} + \epsilon_{yy} - 2 \epsilon_{zz}), \\
        R_{\epsilon} &= \frac{\sqrt{3}}{2} b_v (\epsilon_{xx} - \epsilon_{yy}) - i d_v \epsilon_{xy}, \\
        S_{\epsilon} &= -d_v (\epsilon_{xz} - i \epsilon_{yz}),
    \end{aligned}
\end{equation}
where \(\epsilon_{ij}\) are the components of the strain tensor in three-dimensional space. The constants \(a_v\), \(b_v\), and \(d_v\) are the hydrostatic, uniaxial, and shear deformation potentials, respectively, which describe how the band energy shifts with changes in the strain tensor \cite{Luttinger, willatzen2009kp}. 

The unstrained ground state of a germanium hole spin is the HH state, but strain can significantly hybridize the HH and LH states, altering the HH-LH energy gap \(\Delta_{\text{LH}}\) as follows \cite{mauro2024strain}:
\begin{eqnarray}
    \label{delta_LH}
    \Delta_{\text{LH}} \approx \frac{2\pi^2 \hbar^2 \gamma_2}{m_0 L_w^2} + b_v \left( \langle \epsilon_{xx} \rangle + \langle \epsilon_{yy} \rangle - 2 \langle \epsilon_{zz} \rangle \right).
\end{eqnarray}
where $\gamma_2$ is a Luttinger parameter of Ge and $L_w$ is the thickness of the Ge layer.

The resulting strain-induced corrections to the g-tensor components are:
\begin{eqnarray}
    \label{g-correction}
    \delta g_{xx} = \delta g_{yy} &&= \frac{6 b_v \kappa}{\Delta_{\text{LH}}} \left( \langle \epsilon_{yy} \rangle - \langle \epsilon_{xx} \rangle \right), \\
    \delta g_{zy} &&= -\frac{4 \sqrt{3} \kappa d_v}{\Delta_{\text{LH}}} \langle \epsilon_{yz} \rangle, \\
    \delta g_{zx} &&= -\frac{4 \sqrt{3} \kappa d_v}{\Delta_{\text{LH}}} \langle \epsilon_{xz} \rangle, \\
    \delta g_{xy} = -\delta g_{yx} &&= \frac{4 \sqrt{3} d_v \kappa}{\Delta_{\text{LH}}} \langle \epsilon_{xy} \rangle.
\end{eqnarray}
where \(\langle \epsilon_{ij} \rangle\) denotes the average strain component \(\epsilon_{ij}\) weighted by hole probability density over Germanium layer and $\kappa=3.41$ is the isotropic Zeeman parameter of Germanium~\cite{lawaetz1971valence}. Hence, to calculate the correction to the effective g-tensor of the quantum dot, we just take the expectation value with respect to the hole wavefunction of the dot.

\subsection{Hole Wavefunction Determination}

In order to find the probability distribution of the holes in our device, we simulate the relevant portion using MaSQE, a custom Schrödinger-Poisson solver \cite{MaSQE}. In order to achieve our desired accuracy for the hole's wavefunction we used a multi-domain/multi model simulation with a composite overlapping mesh discretization. Our mesh consists of three components, a global domain for calculation of the hole density in source and drain reservoirs, as well as for simulating our sensing dot, and two higher resolution subdomains in order to simulate both quantum dot charge densities. One hole is always present in each quantum dot subdomain, but the charge in the global domain is allowed to vary. Gate biases are used such that there is no overlap between the global domain and quantum dot subdomain charge distributions. The material parameters we need for this simulation are the following: dielectric constant, effective mass, and the band shift between materials. 

To find the appropriate gate biases to use we compare our simulation with experimental data. The voltages in our simulation are not true voltages, but rather there is a constant offset that is determined by the quantum dot's Fermi level. Therefore, we first set our voltages to coincide with those that were found experimentally, and then we set a voltage offset such that the Fermi levels of our quantum dots allow one hole in each dot. Once this has been determined, we then tune the gate voltages in order to move the dots into regions of large g-factor correction. Of course, changing gate biases also affects the Fermi level of the dots, so we first change the barrier gate voltages in order to move the dots either further away or closer to the middle barrier gate, and then we adjust the plunger gate voltage in order to bring the quantum dot's Fermi level to the appropriate value. Tuning the device in this way can take significant computation time so we tune the device in this way for five different configurations, and then perform a polynomial fit for each gate voltage as a function of the middle barrier gate voltage. This ensures that our device is tuned for each value of middle barrier gate voltage in our simulation. In order to speed up computation of the expectation value of our hole's wavefunction in the g-tensor distribution field we then fit the hole wavefunction to a multivariate Gaussian for each configuration. The values we use is summarized in Table \ref{tab:MaSQE Parameters}.
\begin{table}[t]
    \centering
    \caption{Material Parameters for Wavefunction Simulation}
    \begin{tabular}{l|c|c|c|c}
        \toprule
        Material & $\varepsilon_r$ & $\parallel m^*_h$  & $\perp m^*_h$ & Bandshift (eV) \\
        \midrule
        Ge                       & 15.8  & 0.055 & 0.204 & -0.137\\
        Ge\(_{0.8}\)Si\(_{0.2}\) & 14.99 & 0.414 & 1.527 & 0.0 \\
        Al\(_2\)O\(_3\)          & 7.3   & 6.3   & 0.36  & 1.0 \\
        \bottomrule
    \end{tabular}
    \label{tab:MaSQE Parameters}
\end{table}

\section{Results}

\subsection{Displacement}

\begin{figure*}
    \centering
    \includegraphics[width=1\linewidth]{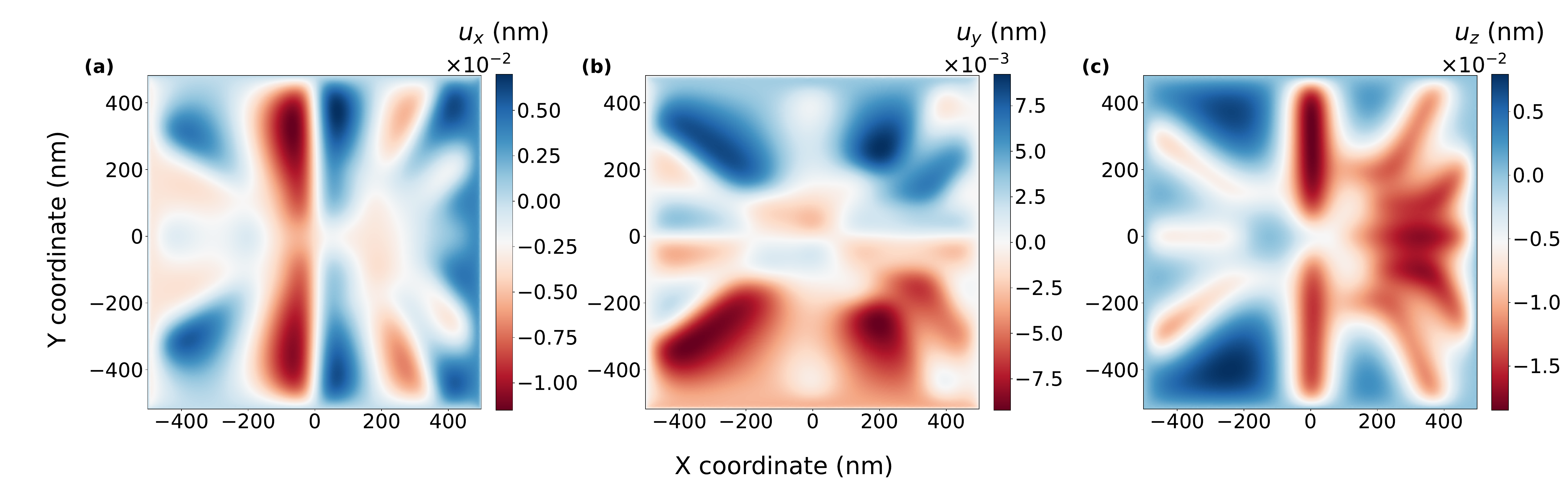}
    \caption{Simulated distribution of the three components of the displacement vector as the device is cooled from the unstrained temperature \(T_0 = 300\,\text{K}\) to the operating temperature \(T = 20\,\text{K}\). The displacement vector at each point indicates how much it moves when the device is cooled down. The strain tensor can be written as a function of only the displacement vector, so we will use the displacement values to calculate strain. The graph shows the distribution on a cross-section through the middle of the germanium layer. (a) x-component of the displacement. (b) y-component of the displacement. (c) z-component of the displacement.}
    \label{displacement_components}
\end{figure*}

In Eq.~\ref{Elasticity}, what we solve for is the displacement vector, which consists of three components at each point in space. From the displacement vector, the strain tensor can be derived by taking its gradient, as outlined in Eq.~\ref{eps_ik}. Using the assumptions, solver settings, parameters, and boundary conditions detailed in Appendix~\ref{FEM Details}, we computed the distribution of the displacement vector components. Notably, we obtained a full 3D solution within the germanium well. While the displacement vector may vary with height, this variation is not the primary focus of our analysis, as the displacement vector serves as an intermediate step to determine the strain. Therefore, we will not delve into the detailed impact of height here. Instead, we provide a brief visualization of the displacement vector component distribution in a horizontal cross-section at \(z = 50\,\text{nm}\), located 8\,nm below the top surface of the germanium well, as shown in Fig.~\ref{displacement_components}. A more detailed discussion of the strain variation with height will be presented in a later section.

\subsection{Strain Distribution}
\begin{figure*}
    \centering
    \includegraphics[width=1\linewidth]{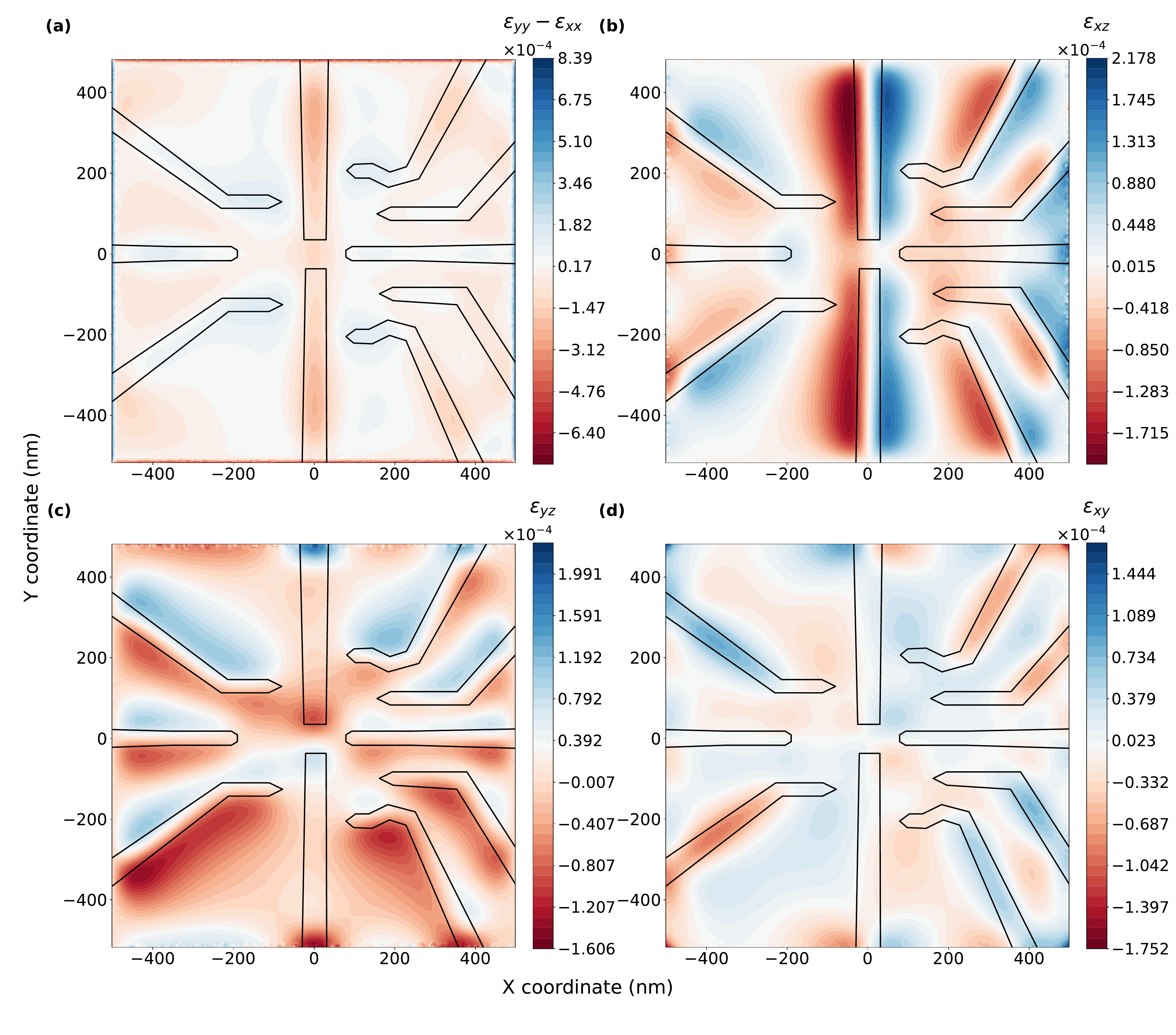}
    \caption{Simulated distribution of selected components of the strain tensor as the device is cooled from the unstrained temperature \(T_0 = 300\,\text{K}\) to the operating temperature \(T = 20\,\text{K}\). The graph shows the distribution on a cross-section at the middle of the germanium layer. Dark lines indicate the locations of the electric gates defining the quantum dot. The strain distribution is highly inhomogeneous and reflects the symmetry of the gate structure, suggesting a strong correlation between the strain pattern and gate design. (a) Difference between the \(\epsilon_{yy}\) and \(\epsilon_{xx}\) components of the strain. (b) \(\epsilon_{xz}\) component of the strain. (c) \(\epsilon_{yz}\) component of the strain. (d) \(\epsilon_{xy}\) component of the strain.}
    \label{Strain}
\end{figure*}
\begin{figure}
    \centering
    \includegraphics[width=1\linewidth]{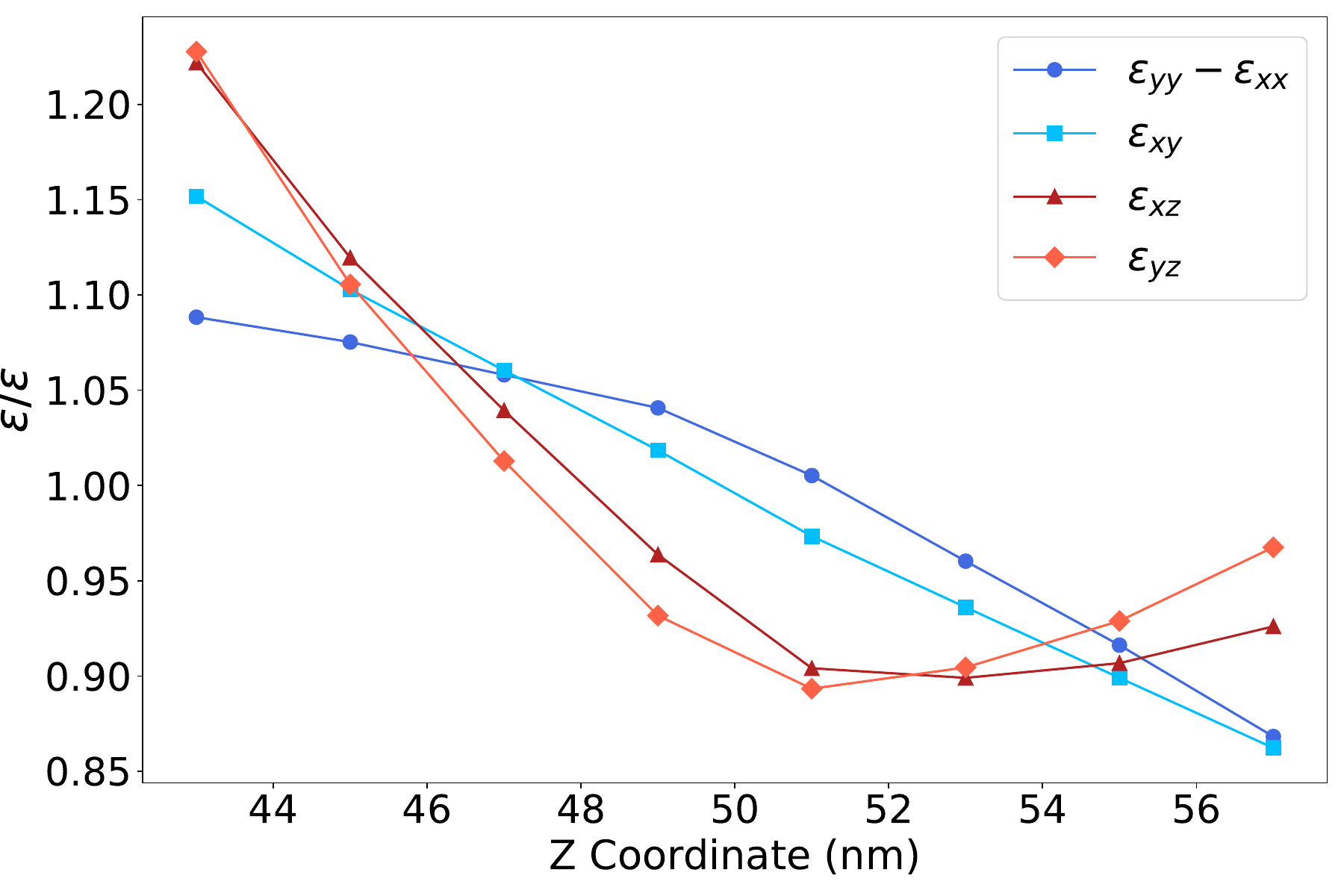}
    \caption{Average magnitude of strain in each of eight horizontal subsets of the germanium well $\epsilon$, divided by the overall average magnitude of that strain component in the entire germanium well $\bar{\epsilon}$. The subsets are equally spaced in the z-direction. The strain is larger at lower z-values, indicating that it builds up as it approaches the gates. The variation in each strain component is about 20-30\% of the overall average, showing that strain varies significantly with depth.}
    \label{average_strain_magnitudes}
\end{figure}
From the displacement vector, we obtained solutions for the strain tensor components \(\epsilon_{yy} - \epsilon_{xx}\), \(\epsilon_{xz}\), \(\epsilon_{yz}\), and \(\epsilon_{xy}\) throughout the entire volume of the germanium well. We plotted the strain distributions at several different z-coordinates and found that the patterns are almost identical across these slices. In Fig.~\ref{Strain}, we present the strain distribution at \(z = 50\,\text{nm}\). We observe that the strain distributions are highly inhomogeneous, forming complex patterns with maximum and minimum strain values occurring within spans of a few tens of nanometers. The strain distribution also mirrors the symmetry of the gate structure, suggesting a strong correlation between the inhomogeneous strain in the germanium quantum dot device and the gate design. This observation is consistent with the findings of \cite{frink2024reducing}, which specifically studied the effect of gate geometry on strain distribution. They examined different gate structures—including a single wire, multiple parallel wires, and a quadruple quantum dot device—and found that gate structure significantly influences strain patterns.

Additionally, we find that the magnitudes of the selected strain components are on the order of \(10^{-4}\), which aligns with the experimental findings of \cite{corley-wiciak2023nanoscale}. They used x-ray mapping to determine strain tensor values in a quantum dot device with a gate structure similar to ours. Our results also correspond with those reported in \cite{abadillo-uriel2023hole}, although their \(\epsilon_{xy}\) values are about a factor of five smaller than ours. This discrepancy may be attributed to differences in gate structure; specifically, their configuration has reflection symmetry with respect to both the x-axis and y-axis, whereas ours has symmetry only with respect to the x-axis.

Despite the similarity in patterns across different z-coordinates, the magnitude of the strain components exhibits slight variations. We divided the germanium well into eight horizontal subsets with equal ranges in the z-direction. In Fig.~\ref{average_strain_magnitudes}, we plot the ratio of the average magnitude of strain in each subset, \(\epsilon\), to the overall average magnitude of strain in the entire germanium well, \(\bar{\epsilon}\). We chose $z=0$ at \(13\,\text{nm}\) below interface between gates and \(\text{Ge}_{0.8}\text{Si}_{0.2}\) substrate, and the $z$ coordinate increases downward. As shown, the strain increases as we get closer to the upper surface of the germanium well (i.e., closer to the gates). This trend can potentially be explained by the fact that, in our device, the strain is primarily caused by the different rates of contraction between the gates and the surrounding materials. The variations are about 30\% of \(\bar{\epsilon}\), indicating that while the strains at all depths within the germanium well are roughly of the same order of magnitude, the variation is significant enough to necessitate considering the charge density distribution in the z-direction when calculating the strain correction to the g-tensor of a quantum dot.

\subsection{LH-HH Band Gap}

\begin{figure}
    \centering
    \includegraphics[width=0.9\linewidth]{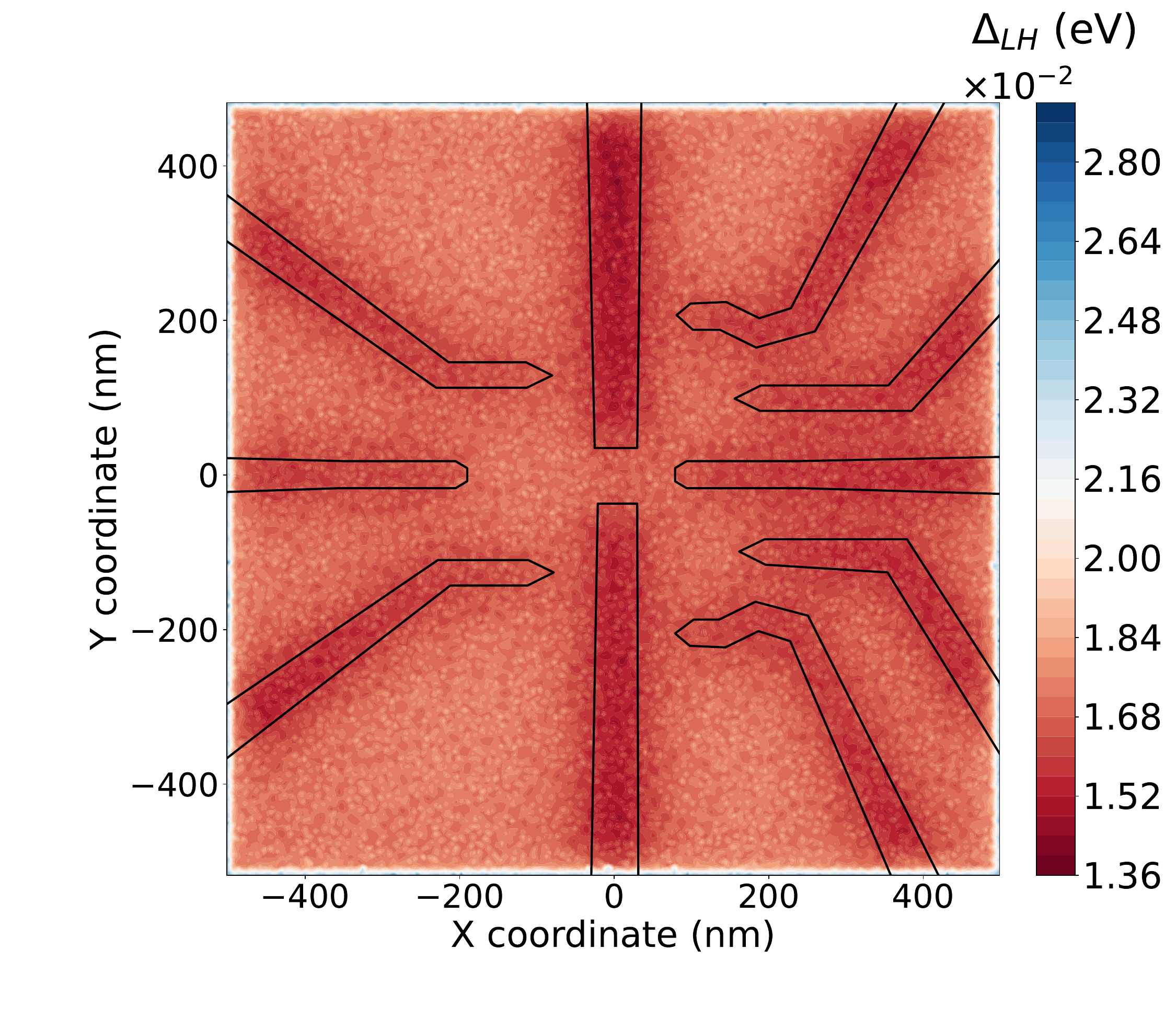}
    \caption{Graph of the HH-LH band gap \(\Delta_{LH}\) at a cross-section in the middle of the germanium layer of our device. The x and y coordinates correspond to the spatial extent in the x and y directions of the cross-section. It can be observed that the strain term \(b_v \left( \langle \epsilon_{xx} \rangle + \langle \epsilon_{yy} \rangle - 2 \langle \epsilon_{zz} \rangle \right)\) is small compared to the constant term in \(\Delta_{LH}\). }
    \label{Band Gap}
\end{figure}

According to Eq.~\ref{g-correction}, the strain correction to the g-tensor components depends on the energy gap between the light-hole and heavy-hole bands, as defined by Eq.~\ref{delta_LH}. Therefore, before calculating the g-tensor correction, it is essential to determine the distribution of this band gap. Figure~\ref{Band Gap} illustrates the band gap distribution in a horizontal cross-section at \(z = 50\,\text{nm}\).
\subsection{g-Tensor Correction Distribution}

\begin{figure*}
    \centering
    \includegraphics[width=1\linewidth]{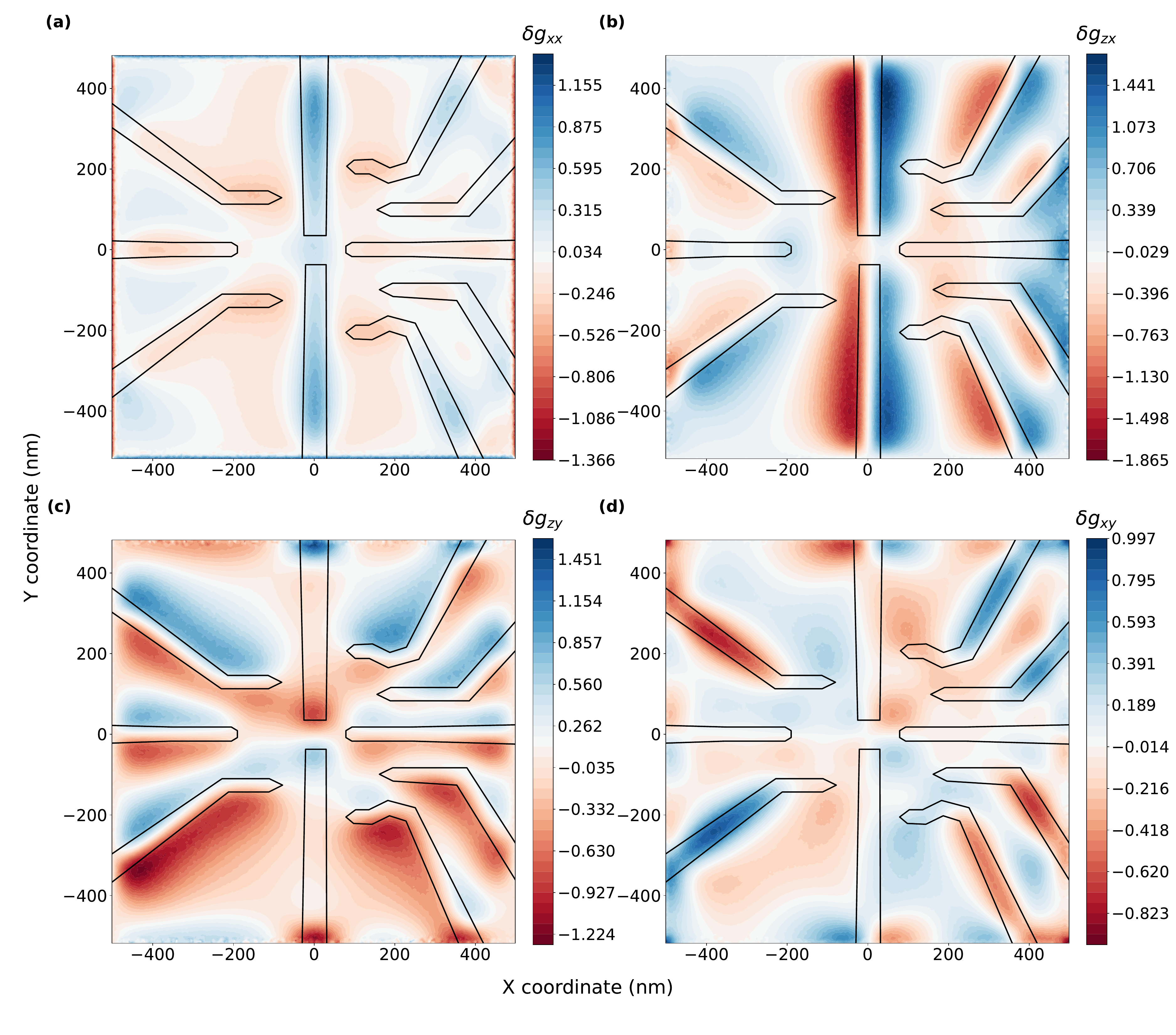}
    \caption{Simulated distribution of selected components of the g-tensor correction as the device is cooled from the unstrained temperature \(T_0 = 300\,\text{K}\) to the operating temperature \(T = 20\,\text{K}\). The graph shows the distribution on a cross-section through the middle of the germanium layer. Dark lines indicate the locations of the electric gates defining the quantum dot. The g-tensor correction is calculated using the strain distribution (Fig. \ref{Strain}) and the LH-HH band gap (Fig. \ref{Band Gap}) with Eq.\ref{g-correction}. Since \(\Delta_{LH}\) is nearly constant across most regions, the inhomogeneity of the g-tensor components largely reflects the distribution of the strain tensor, as they are related by a factor treating \(\Delta_{LH}\) as constant. (a) xx component of g-tensor correction (b) zx component of g-tensor correction (c) zy component of g-tensor correction (d) xy component of g-tensor correction}
    \label{g-correction graph}
\end{figure*}

Using the strain and LH-HH band gap distributions, we calculate the correction to the g-tensor components using Eq.~\ref{g-correction}. We obtained the full 3D solution across the entire germanium well, and the distribution of the g-tensor correction for each component at \(z = 50\,\text{nm}\) is shown in Fig.~\ref{g-correction graph}. We observe that the g-tensor correction follows a pattern similar to that of the strain tensor, as evidenced by the resemblance in their contour lines. The band gap does introduce slight irregularities in the g-tensor correction contours, though its overall effect is minimal. 

It is important to note that the g-tensor distribution graph serves as a visualization of how the g-factor would vary if a hypothetical hole were positioned at each point. Unlike the strain distribution, which directly represents the physical strain at each location, the g-tensor correction must be averaged over the probability density distribution of the hole to yield a physically meaningful value. Therefore, the graph itself does not have direct physical significance but provides insight into the spatial variation of the g-tensor correction.

\subsection{Hole Probability Density Distribution}

Using the MaSQE quantum simulation framework, we systematically investigated the hole probability density distribution in germanium quantum dots under varying gate voltage configurations. For each voltage setting, the spatial distribution of holes was found to exhibit a three-dimensional Gaussian profile, as confirmed by rigorous fitting procedures. By extracting the covariance matrix and mean position vector from the simulated data, we reconstructed the full 3D hole density distribution at arbitrary spatial coordinates within the germanium well. This parametric representation enables efficient calculation of quantum mechanical observables across the confinement potential.  

\begin{figure}
    \centering
    \includegraphics[width=1\linewidth]{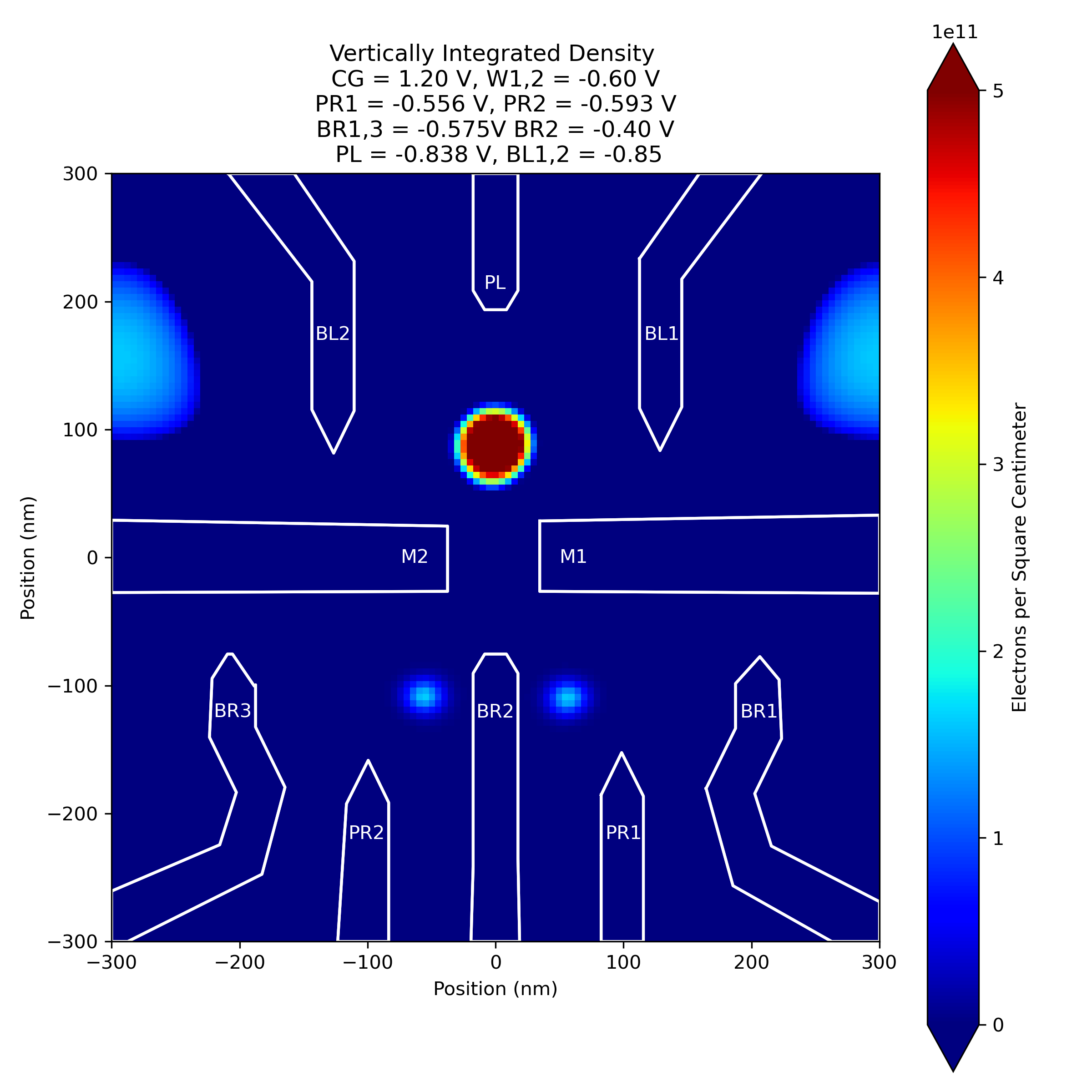}
    \caption{Horizontal planar cross section view of the extent of the hole density distribution under a particular gate voltage setting outlined in the upper captions of the picture. The two bright spots near BR2 gate are the quantum dot used for qubit (we call the left and the right dot), and the bright red spot on the upper half of the picture is the sensing dot.}
    \label{Hole Density Planar}
\end{figure}

The xy-plane cross section of a representative hole density distribution is illustrated in Fig.~\ref{Hole Density Planar}. To generate this profile, we analyzed all density values along the z-axis at fixed \((x,y)\) coordinates within the quantum dot's spatial extent. The excellent agreement between the simulated data (points) and Gaussian fit (curve) demonstrates the validity of our 3D modeling approach. Equivalent Gaussian characteristics were observed in the x- and y-direction projections.

For a visualization of the horizontal planar distribution of the extent of the quantum dot region see Fig. \ref{Hole Density Planar}

\begin{figure}
    \centering
    \includegraphics[width=1\linewidth]{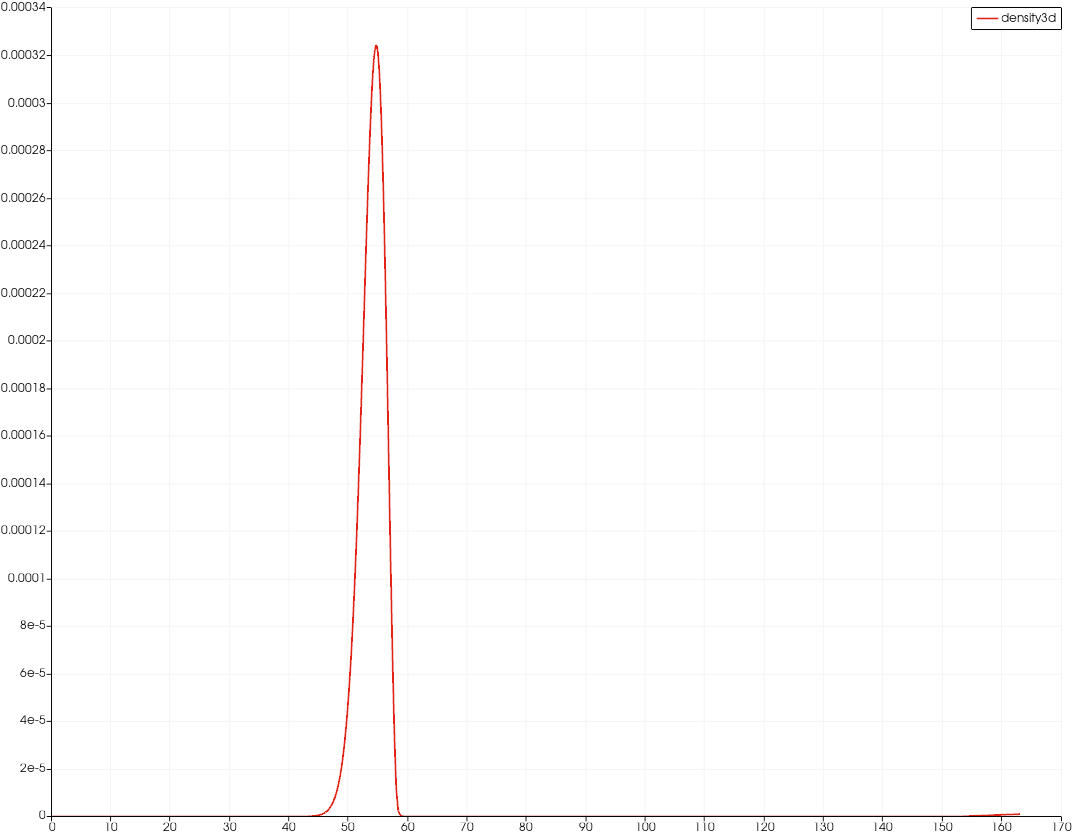}
    \caption{Cross-sectional view of hole probability density along the z-direction at fixed \((x,y)\) coordinates within the quantum dot. Simulated data (circles) shows close agreement with the fitted Gaussian profile (solid curve), validating the 3D normal distribution model. Analogous Gaussian behavior is observed in orthogonal spatial dimensions (x/y directions).}
    \label{Hole_Density_Distribution}
\end{figure}

\subsection{Modulation of g-Tensor Components}

We have shown that the probability density distribution of each quantum dot under each voltage setting can be modeled by a three-dimensional Gaussian function. By averaging the g-tensor correction components weighted by the Gaussian distribution at each voltage setting, we obtain the modulation of each g-tensor correction component.

\begin{figure}
    \centering
    \includegraphics[width=1\linewidth]{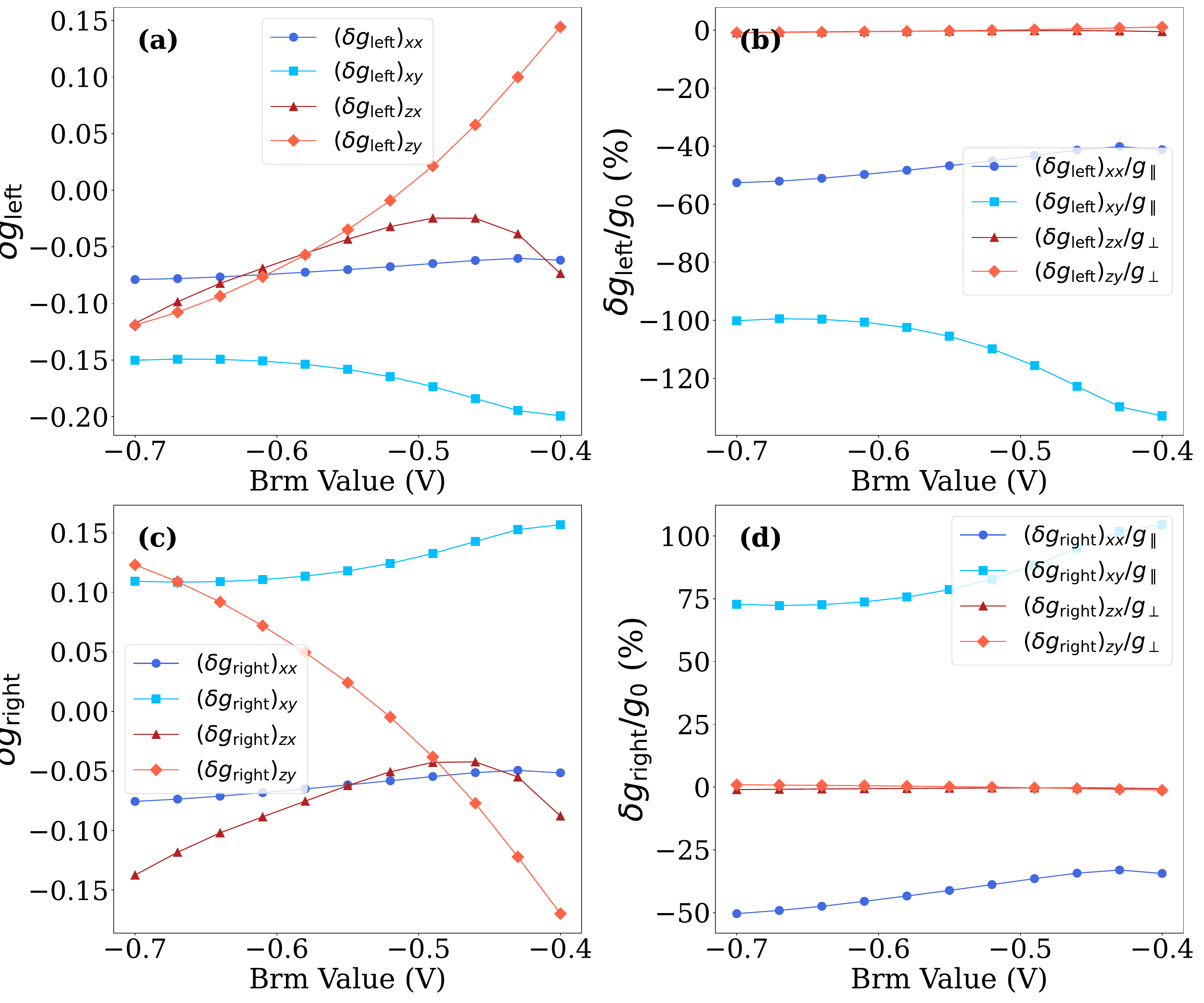}
    \caption{Modulation of strain-induced corrections to the g-tensor components with gate voltage, obtained by averaging the corresponding g-tensor correction components weighted by the hole probability density distribution at each gate voltage setting. (a) Modulation of components of the g-tensor correction of the left dot \( \delta g_{\text{left}} \). (b) Percentage change of the g-tensor correction components of the left dot. \( g_0 \) represents the unstrained g-tensor component value; for in-plane components \( (\delta g)_{xx} \) and \( (\delta g)_{xy} \), \( g_0 = g_\parallel \), and for out-of-plane components \( (\delta g)_{zx} \) and \( (\delta g)_{zy} \), \( g_0 = g_\perp \). (c) Modulation of components of the g-tensor correction of the right dot \( \delta g_{\text{right}} \). (d) Percentage change of the g-tensor correction components of the right dot; the unstrained value \( g_0 \) is defined the same way as for the left dot.}
    \label{component modulation}
\end{figure}

In Fig.~\ref{component modulation}, we plot the modulation of the corrections to the g-tensor components for the left and right dots, respectively. We also display the percentage change of each component with respect to the unstrained value. We take the unstrained out-of-plane g-tensor component to be \( g_\perp = 13.5 \) and the in-plane component to be \( g_\parallel = 0.15 \) \cite{abadillo-uriel2023hole}.

We discover that the modulation of the out-of-plane \( zx \) and \( zy \) components (\( (\delta g_{\text{left}})_{zx} \), \( (\delta g_{\text{left}})_{zy} \), \( (\delta g_{\text{right}})_{zx} \), \( (\delta g_{\text{right}})_{zy} \)) is less than 1\% of \( g_\parallel \), meaning that they are negligible compared to unstrained values. In contrast, the corrections of the in-plane components are on the same order of magnitude as the unstrained values. Specifically, the modulation of the correction to the \( xx \) component (\( (\delta g_{\text{left}})_{xx} \), \( (\delta g_{\text{right}})_{xx} \)) is about 10\% of the unstrained value, and approximately 30\% and 25\% modulation is observed for the \( xy \) component (\( (\delta g_{\text{left}})_{xy} \) and \( (\delta g_{\text{right}})_{xy} \) respectively). Additionally, we find that the modulation of the \( xy \) and \( zy \) components is roughly antisymmetric between the left and right dots, while the modulation of the \( xx \) and \( zx \) components is roughly symmetric. 

\section{Effective g-Factor Calculation}

Quantum dot devices with gate structures similar to ours enable quantum dots to form in two separate regions, each with its own g-factor correction. To calculate the effective g-factor of a single dot, we use the Zeeman energy under a magnetic field, given by 
\begin{eqnarray}
\Delta E_z = \mu_B \vb{g} \vdot \va{B}
\end{eqnarray}
where \(\vb{g}\) is a \(3 \times 3\) tensor and \(\va{B}\) is a 3-vector. The effective g-factor is defined as: 
\begin{eqnarray}
\va{g}^* = \frac{\vb{g} \vdot \va{B}}{B}
\end{eqnarray}
The strain-induced correction to the effective g-factor is then:
\begin{eqnarray}
\label{effective g-correction}
\delta \va{g}^* = \frac{\delta \vb{g} \vdot \va{B}}{B}
\end{eqnarray}
Here, \(B\) is the magnitude of the magnetic field, ensuring that the product \(\va{g}^* B\) produces the correct magnetic field response. Restricting the magnetic field to the in-plane direction and expanding Eq.~\eqref{effective g-correction} in terms of the g-tensor correction components, we find:
\begin{eqnarray}
\delta \va{g}^* =
\begin{bmatrix}
    \delta g_{xx} \cos\phi + \delta g_{xy} \sin\phi \\
    -\delta g_{xy} \cos\phi + \delta g_{xx} \sin\phi \\
    \delta g_{zx} \cos\phi + \delta g_{zy} \sin\phi
\end{bmatrix}
\end{eqnarray}
where \(\phi\) is the azimuthal angle of the magnetic field \(\va{B}\). Note how the out-of-plane g-tensor correction components mix into the in-plane effective g-tensor calculation. This means the modulation of  strain-induced correction to out-of-plane components might magnify the in-plane effective g-factor even if the in-plane g-tensor correction components do not change as much.

\begin{figure*}
    \centering
    \includegraphics[width=1\linewidth]{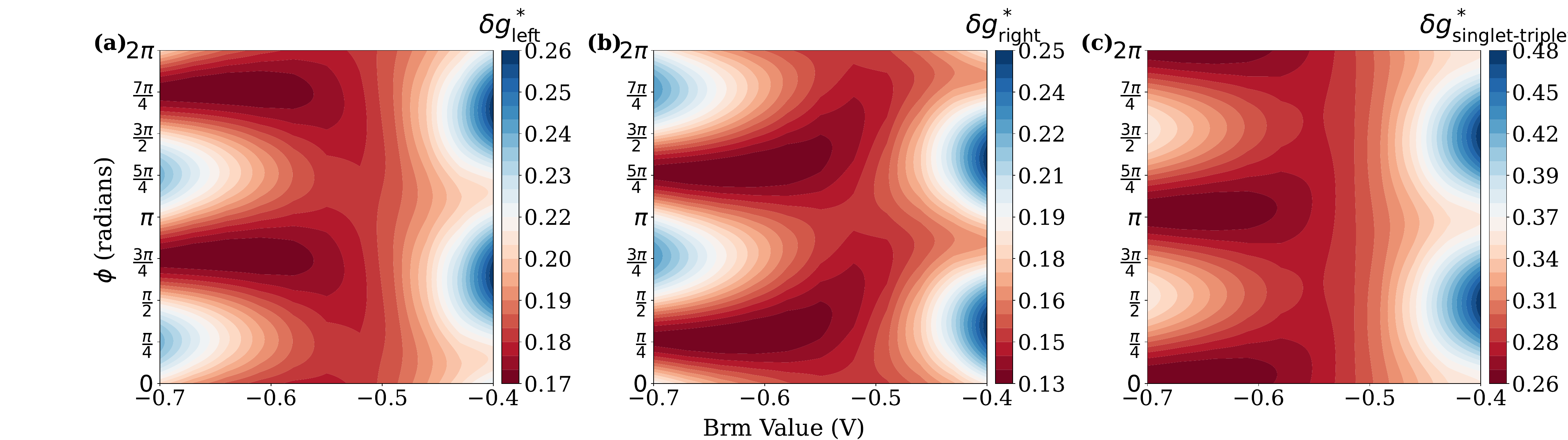}
    \caption{Contour plots of the effective g-factor magnitude, with Brm gate voltage values on the $x$-axis and magnetic field azimuthal angles on the $y$-axis. Contour lines indicate constant effective g-factor magnitudes, revealing how specific combinations of magnetic field orientation and gate voltage can mitigate strain effects. (a) Effective g-factor magnitude for the left dot. A 0.2 V change in Brm results in a magnitude shift of approximately 0.09. (b) Effective g-factor magnitude for the right dot. A 0.2 V change in Brm leads to a magnitude shift of about 0.12. (c) Effective g-factor magnitude for the singlet-triplet qubit. A 0.2 V change in Brm combined with a $\frac{\pi}{2}$ rotation of the magnetic field causes the magnitude to change of 0.22}
    \label{g_eff_contour}
\end{figure*}

For a singlet-triplet qubit, $\delta \vb{g}$ is replaced by $\delta \vb{g}_{\text{left}} - \delta \vb{g}_{\text{right}}$, where $\vb{g}_{\text{left}}$ and $\vb{g}_{\text{right}}$ represent the strain-induced corrections to the g-tensors of the left and right quantum dots, respectively. To investigate the interplay between in-plane magnetic field orientation and gate voltage on the effective g-tensor, we construct a contour map with gate voltages on the $x$-axis, the magnetic field's azimuthal angle on the $y$-axis, and the magnitude of the effective g-tensor reflected in the color, as illustrated in Fig.~\ref{g_eff_contour}.

Using an uncorrected in-plane g-factor $g_\parallel = 0.15$, we observe that the effective g-factor for a single dot can vary by over 80\%, while the singlet-triplet qubit's effective g-factor varies by more than 140\%. These substantial variations underscore the critical influence of strain-induced effects under an in-plane magnetic field.

Beyond the magnitude changes, the contour patterns reveal several key insights. First, significant variations in the effective g-factor can be achieved by independently adjusting either the magnetic field orientation or the Brm gate voltage. For instance, in the singlet-triplet case, fixing Brm at $-0.7$ and varying the azimuthal angle from $\frac{3\pi}{2}$ to $\pi$ results in a 73\% change. Conversely, fixing $\phi = \frac{3\pi}{2}$ and increasing Brm from $-0.55$ to $-0.4$ leads to a 133\% change. Second, simultaneous adjustments of the magnetic field orientation and gate voltage along the contour lines can stabilize the effective g-factor, offering a practical approach to minimizing strain effects. Additionally, for certain magnetic field orientations, the effective g-factor of the singlet-triplet qubit exhibits non-monotonic behavior with gate voltage changes. For example, at $\phi = \frac{3\pi}{2}$, $\delta g^*$ initially decreases and then increases, a phenomenon also experimentally observed in \cite{rooney2023gate}.

\section{Conclusion}

This study aimed to understand the gate modulation of the g-tensor observed in previous research, which has been suspected to result from strain effects. To quantitatively analyze this, we simulated the thermal contraction strain distribution in a quantum dot device by solving the thermoelasticity equation using FEM. From the solution, we extracted the strain distribution and the g-tensor correction for each component throughout the entire volume of the germanium well.

Next, we solved Schrödinger’s equation with an electric potential background obtained from Poisson’s equation using a high-order finite difference method. This allowed us to model the hole probability density distribution as a function of gate voltage. At each voltage setting, we discovered that the probability density distribution can be described by a three-dimensional Gaussian function. By taking the weighted average of the g-tensor corrections using the quantum dot's probability density distribution, we calculated the corrected g-tensor components for each dot under various voltage settings.

Our findings indicate that the corrections to the out-of-plane g-tensor components are less than 1\% of the unstrained values. In contrast, for the in-plane components, both the strain-induced corrections and their modulation are on the same order of magnitude as the uncorrected values. Finally, we calculated the effective g-factor for a single dot and a singlet-triplet qubit under different in-plane magnetic field orientations and gate voltages. We found that the combined effect of the magnetic field and gate voltage gives rise to complex patterns in the effective g-factor. Notably, within a 0.3\,V change in gate voltage and a \(\frac{\pi}{2}\) rotation of the in-plane magnetic field, the effective g-factor for a single dot changes by about 80\%, and that of a singlet-triplet qubit changes by 140\% compared to the unstrained in-plane g-factor $g_\parallel = 0.15$. The g-tensor correction patterns with respect to magnetic field and gate voltage also suggest multiple approaches to tune the gate voltage and magnetic field orientation to minimize the strain effects.

This study provides a systematic method for investigating strain effects in germanium quantum dot devices with arbitrary gate structures, operating temperatures, and substrate materials. We demonstrated that the observed gate modulation of the g-factor in recent studies can potentially be attributed to strain effects. We have shown that strain effects are significant when the device operates under an in-plane magnetic field and highlighted strategies to minimize or maximize strain-induced corrections to the effective g-factor. This in-depth study offers important insights into quantum dot device design and qubit control.

This study has several limitations. Firstly, the effect of gate structure on the strain pattern was not extensively explored. Our findings indicate that gate structure has a direct impact on the strain field—for example, the strain is larger near the gates, and the strain field reflects the symmetry of the gate structure. Future studies could simulate strain distributions for various gate structures to investigate how gate materials, thickness, and symmetry affect each strain component.

Additionally, this study focused on the effective g-factor when the qubit operates under an in-plane magnetic field. Due to the large anisotropy of the g-tensor in germanium, even a slight tilt of the magnetic field from the horizontal plane can significantly change the effective g-factor. While we found that the out-of-plane correction components are negligible compared to the unstrained values, above a certain tilt angle, we expect to find a g-factor that is relatively immune to strain effects. Determining this specific angle, and understanding how the interplay between gate voltage, polar angle, and azimuthal angle of the magnetic field affects the effective g-factor, are areas for future investigation.

Lastly, this study simulated strain by assuming a constant thermal expansion coefficient, averaged over different temperatures, and we made the assumption of weak coupling, where strain is assumed to have no thermal effects. The validity of these assumptions could be questioned. Future studies might employ a more comprehensive coupling model between heat and strain or simulate the entire cooling process to examine these factors in detail.

\section{Data Availability Statement}

The simulation code, mesh files, and archived run data supporting this study are openly available on Zenodo~\cite{niu2026gfsimulation}.

\begin{acknowledgments}
Research was sponsored by the Air Force Office of Scientific Research (AFOSR) under award number FA9550-23-1-0710 (at UCLA) and was sponsored by the Army Research Office (ARO) and was accomplished under Grant No.W911NF-23-1-0016. The views and conclusions contained in this document are those of the authors and should not be interpreted as representing the official policies, either expressed or implied, of the Army Research Office (ARO), the Air Force Office Scientific Research (AFOSR), or the U.S. Government. The U.S. Government is authorized to reproduce and distribute reprints for Government purposes notwithstanding any copyright notation herein.
\end{acknowledgments}

\newpage

\bibliography{references_prb}% Produces the bibliography via BibTeX.

@PREAMBLE{
 "\providecommand{\noopsort}[1]{}" 
 # "\providecommand{\singleletter}[1]{#1}%" 
}

@article{Kloeffel,
  author  = {Kloeffel, Christoph and Trif, Mircea and Loss, Daniel},
  title   = {Strong spin-orbit interaction and helical hole states in {Ge/Si} nanowires},
  journal = {Phys. Rev. B},
  volume  = {84},
  pages   = {195314},
  year    = {2011},
  doi     = {10.1103/PhysRevB.84.195314}
}

@article{Scappucci,
  author  = {Scappucci, Giordano and Kloeffel, Christoph and Zwanenburg, Floris A. and Loss, Daniel and Myronov, Maksym and Zhang, Jian-Jun and De Franceschi, Silvano and Katsaros, Georgios and Veldhorst, Menno},
  title   = {The germanium quantum information route},
  journal = {Nat. Rev. Mater.},
  volume  = {6},
  pages   = {926--943},
  year    = {2021},
  doi     = {10.1038/s41578-020-00262-z}
}

@article{watzinger2018germanium,
  author  = {Watzinger, Hannes and Kuku\v{c}ka, Josip and Vuku\v{s}i\'{c}, Lada and Gao, Fei and Wang, Ting and Sch\"{a}ffler, Friedrich and Zhang, Jian-Jun and Katsaros, Georgios},
  title   = {A germanium hole spin qubit},
  journal = {Nat. Commun.},
  volume  = {9},
  pages   = {3902},
  year    = {2018},
  doi     = {10.1038/s41467-018-06418-4}
}

@article{jirovec2021singlet,
  author  = {Jirovec, Daniel and Hofmann, Andrea and Ballabio, Andrea and Mutter, Philipp M. and Tavani, Giulio and Botifoll, Marc and Crippa, Alessandro and others},
  title   = {A singlet-triplet hole spin qubit in planar {Ge}},
  journal = {Nat. Mater.},
  volume  = {20},
  pages   = {1106--1112},
  year    = {2021},
  doi     = {10.1038/s41563-021-01022-2}
}

@article{zajac2018resonantly,
  author  = {Zajac, D. M. and Sigillito, A. J. and Russ, M. and Borjans, F. and Taylor, J. M. and Burkard, G. and Petta, J. R.},
  title   = {Resonantly driven {CNOT} gate for electron spins},
  journal = {Science},
  volume  = {359},
  pages   = {439--442},
  year    = {2018},
  doi     = {10.1126/science.aao5965}
}

@article{petta2005coherent,
  author  = {Petta, J. R. and Johnson, A. C. and Taylor, J. M. and Laird, E. A. and Yacoby, A. and Lukin, M. D. and Marcus, C. M. and Hanson, M. P. and Gossard, A. C.},
  title   = {Coherent manipulation of coupled electron spins in semiconductor quantum dots},
  journal = {Science},
  volume  = {309},
  pages   = {2180--2184},
  year    = {2005},
  doi     = {10.1126/science.1116955}
}

@article{rooney2023gate,
  author  = {Rooney, John and Luo, Zhentao and Stehouwer, Lucas E. A. and Scappucci, Giordano and Veldhorst, Menno and Jiang, Hong-Wen},
  title   = {Gate modulation of the hole singlet-triplet qubit frequency in germanium},
  journal = {npj Quantum Inf.},
  volume  = {11},
  pages   = {15},
  year    = {2025},
  doi     = {10.1038/s41534-024-00953-3}
}

@article{hendrickx2024sweet,
  author  = {Hendrickx, N. W. and Massai, L. and Mergenthaler, M. and Schupp, F. J. and Paredes, S. and Bedell, S. W. and Salis, G. and Fuhrer, A.},
  title   = {Sweet-spot operation of a germanium hole spin qubit with highly anisotropic noise sensitivity},
  journal = {Nat. Mater.},
  volume  = {23},
  pages   = {920--927},
  year    = {2024},
  doi     = {10.1038/s41563-024-01857-5}
}

@article{chekhovich2015suppression,
  author  = {Chekhovich, E. A. and Hopkinson, M. and Skolnick, M. S. and Tartakovskii, A. I.},
  title   = {Suppression of nuclear spin bath fluctuations in self-assembled quantum dots induced by inhomogeneous strain},
  journal = {Nat. Commun.},
  volume  = {6},
  pages   = {6348},
  year    = {2015},
  doi     = {10.1038/ncomms7348}
}

@article{abadillo-uriel2023hole,
  author  = {Abadillo-Uriel, Jos\'e Carlos and Rodr\'iguez-Mena, Esteban A. and Martinez, Biel and Niquet, Yann-Michel},
  title   = {Hole-spin driving by strain-induced spin-orbit interactions},
  journal = {Phys. Rev. Lett.},
  volume  = {131},
  pages   = {097002},
  year    = {2023},
  doi     = {10.1103/PhysRevLett.131.097002}
}

@article{michal2021longitudinal,
  author  = {Michal, Vincent P. and Venitucci, Benjamin and Niquet, Yann-Michel},
  title   = {Longitudinal and transverse electric field manipulation of hole spin-orbit qubits in one-dimensional channels},
  journal = {Phys. Rev. B},
  volume  = {103},
  pages   = {045305},
  year    = {2021},
  doi     = {10.1103/PhysRevB.103.045305}
}

@misc{taboga2021properties,
  author       = {Taboga, Marco},
  title        = {Properties of eigenvalues and eigenvectors},
  howpublished = {Lectures on Matrix Algebra (StatLect),
                  \url{https://www.statlect.com/matrix-algebra/properties-of-eigenvalues-and-eigenvectors}},
  year         = {2021}
}

@book{landau1986theory,
  author    = {Landau, L. D. and Lifshitz, E. M. and Kosevich, A. M. and Pitaevskii, L. P.},
  title     = {Theory of {E}lasticity},
  series    = {Course of {T}heoretical {P}hysics},
  volume    = {7},
  edition   = {3},
  publisher = {Butterworth-Heinemann},
  address   = {Oxford},
  year      = {1986}
}

@misc{mauro2024strain,
  author        = {Mauro, Lorenzo and Rodr\'iguez-Mena, Esteban A. and Martinez, Biel and Niquet, Yann-Michel},
  title         = {Strain engineering in {Ge/GeSi} spin qubits heterostructures},
  year          = {2024},
  eprint        = {2407.19854},
  archivePrefix = {arXiv},
  primaryClass  = {cond-mat.mes-hall},
  doi           = {10.48550/arXiv.2407.19854}
}

@article{Luttinger,
  author  = {Luttinger, J. M. and Kohn, W.},
  title   = {Motion of electrons and holes in perturbed periodic fields},
  journal = {Phys. Rev.},
  volume  = {97},
  pages   = {869--883},
  year    = {1955},
  doi     = {10.1103/PhysRev.97.869}
}

@book{willatzen2009kp,
  author    = {Willatzen, Morten and Lew Yan Voon, Lok C.},
  title     = {The {$\mathbf{k}\cdot\mathbf{p}$} {M}ethod: {E}lectronic {P}roperties of {S}emiconductors},
  publisher = {Springer},
  address   = {Berlin, Heidelberg},
  year      = {2009},
  doi       = {10.1007/978-3-540-92872-0}
}

@article{corley-wiciak2023nanoscale,
  author  = {Corley-Wiciak, Cedric and Richter, Carsten and Zoellner, Marvin H. and Zaitsev, Ignatii and Manganelli, Costanza L. and Zatterin, Edoardo and Sch\"ulli, Tobias U. and others},
  title   = {Nanoscale mapping of the {3D} strain tensor in a germanium quantum well hosting a functional spin qubit device},
  journal = {ACS Appl. Mater. Interfaces},
  volume  = {15},
  pages   = {3119--3130},
  year    = {2023},
  doi     = {10.1021/acsami.2c17395}
}

@book{bleyer2018numericaltours,
  author    = {Bleyer, Jeremy},
  title     = {Numerical {T}ours of {C}omputational {M}echanics with {FE}ni{CS}},
  publisher = {Zenodo},
  year      = {2018},
  doi       = {10.5281/zenodo.1287832}
}

@article{Saad,
  author  = {Saad, Youcef and Schultz, Martin H.},
  title   = {{GMRES}: A generalized minimal residual algorithm for solving nonsymmetric linear systems},
  journal = {SIAM J. Sci. Stat. Comput.},
  volume  = {7},
  pages   = {856--869},
  year    = {1986},
  doi     = {10.1137/0907058}
}

@article{xu2017algebraic,
  author  = {Xu, Jinchao and Zikatanov, Ludmil},
  title   = {Algebraic multigrid methods},
  journal = {Acta Numer.},
  volume  = {26},
  pages   = {591--721},
  year    = {2017},
  doi     = {10.1017/S0962492917000083}
}

@misc{frink2024reducing,
  author        = {Frink, Collin C. D. and Oh, Talise and Joseph, E. S. and Losert, Merritt P. and MacQuarrie, E. R. and Woods, Benjamin D. and Eriksson, M. A. and Friesen, Mark},
  title         = {Reducing strain fluctuations in quantum dot devices by gate-layer stacking},
  year          = {2025},
  eprint        = {2312.09235},
  archivePrefix = {arXiv},
  primaryClass  = {cond-mat.mes-hall}
}

@article{slack1975thermal,
  author  = {Slack, G. A. and Bartram, S. F.},
  title   = {Thermal expansion of some diamondlike crystals},
  journal = {J. Appl. Phys.},
  volume  = {46},
  pages   = {89--98},
  year    = {1975},
  doi     = {10.1063/1.321373}
}

@book{levinshtein2001properties,
  author    = {Levinshtein, M. E. and Rumyantsev, S. L. and Shur, M. S.},
  title     = {Properties of {A}dvanced {S}emiconductor {M}aterials: {GaN}, {AlN}, {InN}, {BN}, {SiC}, {SiGe}},
  publisher = {John Wiley \& Sons},
  address   = {New York},
  year      = {2001}
}

@article{schauer1965thermal,
  author  = {Schauer, A.},
  title   = {Thermal expansion, {G}r\"uneisen parameter, and temperature dependence of lattice vibration frequencies of aluminum oxide},
  journal = {Can. J. Phys.},
  volume  = {43},
  pages   = {523--531},
  year    = {1965},
  doi     = {10.1139/p65-049}
}

@article{Alnaes2015FEniCS,
  author  = {Aln\ae{}s, M. S. and Blechta, J. and Hake, J. and Johansson, A. and Kehlet, B. and Logg, A. and Richardson, C. and Ring, J. and Rognes, M. E. and Wells, G. N.},
  title   = {The {FE}ni{CS} project version 1.5},
  journal = {Arch. Numer. Softw.},
  volume  = {3},
  pages   = {9--23},
  year    = {2015},
  doi     = {10.11588/ans.2015.100.20553}
}

@article{Burkard2023SpinQubits,
  author  = {Burkard, Guido and Ladd, Thaddeus D. and Pan, Andrew and Nichol, John M. and Petta, Jason R.},
  title   = {Semiconductor spin qubits},
  journal = {Rev. Mod. Phys.},
  volume  = {95},
  pages   = {025003},
  year    = {2023},
  doi     = {10.1103/RevModPhys.95.025003}
}

@article{MaSQE,
  author  = {Anderson, Christopher R.},
  title   = {Efficient solution of the {S}chr\"odinger--{P}oisson equations in layered semiconductor devices},
  journal = {J. Comput. Phys.},
  volume  = {228},
  pages   = {4745--4756},
  year    = {2009},
  doi     = {10.1016/j.jcp.2009.03.037}
}

@article{lawaetz1971valence,
  author  = {Lawaetz, Peter},
  title   = {Valence-band parameters in cubic semiconductors},
  journal = {Phys. Rev. B},
  volume  = {4},
  pages   = {3460--3467},
  year    = {1971},
  doi     = {10.1103/PhysRevB.4.3460}
}

@misc{niu2026gfsimulation,
  author = {Niu, Mu and Culver, Adrian and Bryan, Johnathan and Anderson, Chris and Gyure, Mark and Jiang, HongWen},
  title  = {Quantum Dot g-Factor Simulation with {FEniCS}},
  year   = {2026},
  note   = {Version 1.0.0, [Software], Zenodo},
  doi    = {10.5281/zenodo.22695247}
}

\newpage

\appendix

\section{Thermal Elasticity Theory} \label{Thermal Elasticity}

\subsection{Linear Elasticity Theory}

\subsubsection{The Strain Tensor}

Strain is a measure that describes the extent to which a material is stretched or compressed. Focusing on a single point within the material, the deformation manifests as the displacement of that point, defined by the displacement vector:
\begin{equation}
    u_i = x'_i - x_i
\end{equation}
where \(x_i\) and \(x'_i\) are the locations before and after displacement, respectively. Consider two points initially separated by a small distance \(dl = \sqrt{\sum_i x_i^2}\). If their displacement vectors differ by \(du\), then after displacement, their separation becomes \(dl' = \sqrt{\sum_i x_i'^2} = \sqrt{\sum_i (dx_i + du_i)^2}\). Since displacement depends on the original position in all three directions, we can write:
\[
    du_i = \sum_k \pdv{u_i}{x_k} dx_k
\]
Thus, we can express \(dl'^2\) in terms of variables prior to displacement:
\begin{equation}
    \label{dl'}
    \begin{aligned}
        dl'^2 = & \sum_i \left(dx_i + \sum_k \pdv{u_i}{x_k} dx_k\right)^2 \\
        = & \sum_i dx_i^2 + 2 \sum_{i,k} \pdv{u_i}{x_k} dx_k dx_i + \sum_{k,l} \pdv{u_i}{x_k} \pdv{u_i}{x_l} dx_k dx_l
    \end{aligned}
\end{equation}
The sum expression of the second term is symmetric with respect to indices \(i\) and \(k\). By switching indices \(i\) and \(l\) in the third term, equation \ref{dl'} becomes:
\begin{equation}
    \begin{aligned}
        dl'^2 = & dl^2 + \sum_{i,k} \left(\pdv{u_i}{x_k} dx_k dx_i + \pdv{u_k}{x_i} dx_i dx_k\right) \\
        & + \sum_{i,k} \pdv{u_l}{x_k} \pdv{u_l}{x_i} dx_k dx_i \\
        = & dl^2 + \sum_{i,k} 2\epsilon_{ik} dx_i dx_k
    \end{aligned}
\end{equation}
where the strain tensor is defined as:
\begin{equation}
    \label{eps_ik}
    \epsilon_{ik} = \frac{1}{2} \left(\pdv{u_i}{x_k} + \pdv{u_k}{x_i} + \pdv{u_l}{x_i} \pdv{u_l}{x_k}\right)
    \approx \frac{1}{2} \left(\pdv{u_i}{x_k} + \pdv{u_k}{x_i}\right)
\end{equation}
This approximation holds when we assume the strain is small and keep only first order terms, which by definition of strain, requires small displacement for our geometry. It is possible to experience significant displacement even with minor internal deformation. For instance, a thin, long rod may display large end displacements even under minor strain. However, in a 3D body where no dimension is significantly smaller than the others, such displacement remains minimal. This scenario applies to our quantum dot device geometry; thus, we will use this approximated equation throughout the paper.

From equation \ref{eps_ik}, we observe that the strain tensor is symmetric and therefore diagonalizable at any point. Consequently, the expression for \(dl'^2\) becomes decoupled in the directions of the three eigenvectors of the strain tensor:
\begin{equation}
    dl'^2 = \sum_i (1 + 2\epsilon^{(i)}) dx_i^2
\end{equation}
Here, \(\epsilon^{(i)}\) denotes the \(i^{th}\) eigenvalue, and the relative extension along the \(i^{th}\) axis is approximately \(\epsilon^{(i)}\), given by \(\sqrt{1 + 2\epsilon^{(i)}} - 1\). This leads to a change in volume expressed as:
\begin{equation}
    dV' = dV \prod_i (\epsilon^{(i)} + 1) \approx dV (1 + \sum_i \epsilon^{(i)})
\end{equation}
where higher order terms are neglected in the final approximation. Since the sum of the eigenvalues of a matrix equals its trace \cite{taboga2021properties}, we find that the trace of the strain tensor dictates the volume change after deformation, as shown by:
\begin{equation}
    \frac{dV' - dV}{dV} = \Tr(\epsilon) = \sum_i \epsilon_{ii}
\end{equation}
For a traceless strain tensor, the volume remains invariant during deformation (although the shape may change), which we term pure shear. Conversely, if the strain tensor equals a constant times the identity matrix, the shape remains invariant during deformation (while the volume may change), termed hydrostatic deformation. We can always decompose the strain tensor into its pure shear and hydrostatic components, an expression crucial for later discussions:
\begin{equation}
    \epsilon_{ik} = \left(\epsilon_{ik} - \frac{1}{3} \delta_{ik} \Tr(\epsilon)\right) + \frac{1}{3} \delta_{ik} \Tr(\epsilon)
\end{equation}

\subsubsection{The Stress Tensor}

When a body is deformed from equilibrium, an internal force emerges that tends to restore it to its original state. This force is linked to internal stress. Before we delve into the definition of the stress tensor, let's first examine this force. Consider the total force experienced by a volume \(dV\) within the body, which can be expressed as \(\mathbf{F} dV\), where \(\mathbf{F}\) is the force per unit volume, represented as a vector. According to Newton's third law, internal forces within a material cancel each other out, meaning the total force experienced by the volume must equal the total external force acting upon it. This external force acts through the surface of the volume, necessitating that the volume integral \(\int \mathbf{F} dV\) be expressible as a surface integral. According to the divergence theorem, the volume integral of a vector field is equivalent to a surface integral if the vector field is the divergence of some tensor field. Applying this principle to each component of \(\mathbf{F}\), we conclude that \(\mathbf{F}\) must be the divergence of a rank two tensor, typically called the stress tensor, described by the relationship:
\begin{equation}
    \label{F_i}
    F_i = \sum_k \pdv{\sigma_{ik}}{x_k}
\end{equation}
Consequently, the total force can be expressed as:
\begin{equation}
    \int \mathbf{F} dV = \int \sum_k \pdv{\sigma_{ik}}{x_k} dV = \int \sum_k \sigma_{ik} df_k
\end{equation}
This formula gives the interpretation of \(\sum_k \sigma_{ik} df_k\) as the \(i^{th}\) component of force acting on a surface element \(d\mathbf{f}\). Here, \(\sigma_{ik}\) represents the \(i^{th}\) component of force per area acting on the \(k^{th}\) component of the surface, which is normal to direction \(k\). For instance, \(\sigma_{xx}\) is the x-component of the force per area vector acting on a surface perpendicular to the x-axis (a normal force), while \(\sigma_{xy}\) is the y-component of the force acting on a surface perpendicular to the x-axis (a tangential force).

\subsection{Thermodynamic of Elasticity}

\subsubsection{Work Done by Stress}

We previously derived the relationship between force per unit volume and stress in equation \ref{F_i}. To calculate the work done by stress, we multiply the force by a displacement \(\delta u\):
\begin{equation}
    \label{W}
    \begin{aligned}
        W = & \int_\Omega \sum_{k,i} \pdv{\sigma_{ik}}{x_k} \delta u_i dV \\
        = & \oint_{\partial \Omega} \sum_{k,i} \sigma_{ik}\delta u_i df_k - \int_\Omega \sum_{k,i} \sigma_{ik} \pdv{\delta u_i}{x_k} dV
    \end{aligned}
\end{equation}
The second equality is derived using integration by parts with respect to \(x_k\). We can freely choose the domain of integration \(\Omega\) in Eq.~\eqref{W}. Assuming the object is not deformed at infinity, we may let \(\Omega\) encompass the entire space, causing the surface integral to vanish. Thus,
\begin{equation}
\label{Work done by stress}
    \begin{aligned}
        W = & - \int_\Omega \sum_{k,i} \sigma_{ik} \pdv{\delta u_i}{x_k} dV \\
        = & - \frac{1}{2} \int_\Omega \sum_{k,i} \sigma_{ik} \qty(\pdv{\delta u_i}{x_k} + \pdv{\delta u_k}{x_i}) dV \\
        = & - \frac{1}{2} \int_\Omega \sum_{k,i} \sigma_{ik} \delta \qty(\pdv{u_i}{x_k} + \pdv{u_k}{x_i}) dV \\
        = & \int_\Omega \sum_{k,i} \sigma_{ik} \delta \epsilon_{ik} dV
        = \int_\Omega \vb{\sigma(u)}:\vb{\delta \epsilon(u)} dV
    \end{aligned}
\end{equation}
This yields the expression of work in terms of the double contraction of the stress tensor by the change in the strain tensor. This work must be balanced by the external work done on the surface of the object. Assuming the force per unit area of the external force is \(\mathbf{T}\) and that the infinitesimal strain tensor $\epsilon_{ik}$ is induced by infinitesimal displacement $\vb{v}$, we obtain the following equation:
\begin{equation}
    \label{weak form}
    \forall \vb{v} \in V: \quad \int_\Omega \vb{\sigma(u)}:\vb{\epsilon(v)} d\Omega = \int_{\partial \Omega} \vb{T} \vdot \vb{v} dS
\end{equation}
This is the equation that we will solve using FEM in this study, where $\vb{v}$ is the test function and $V$ is the function space which contains both the test function $\vb{v}$ and the solution $\vb{u}$.

\subsubsection{Free Energy of Deformation}

An infinitesimal change in internal energy \(dU\) is equal to the heat transferred to the system minus the work done by the system. Using Eq.~\eqref{Work done by stress}, we have:
\begin{equation}
dU = TdS + \sum_{i,k} \sigma_{ik} d\epsilon_{ik}
\end{equation}
The free energy \(\mathcal{F}\) is related to the internal energy by \(\mathcal{F} = U - TS\). This gives us its total differential:
\begin{equation}
\begin{aligned}
d\mathcal{F} &= dU - TdS - SdT \\
&= -SdT + \sum_{i,k} \sigma_{ik} d\epsilon_{ik}
\end{aligned}
\end{equation}
Then, under constant temperature, we derive an important differential relation that links the stress tensor to the strain tensor:
\begin{equation}
    \label{sigma}
    \sigma_{ik} = \left(\pdv{\mathcal{F}}{\epsilon_{ik}}\right)_T
\end{equation}

\subsubsection{Expansion of Free Energy}

We aim to expand the free energy up to the second order of the strain tensor, represented as \(\mathcal{F} = \mathcal{F}_0 + \mathcal{F}_1 + \mathcal{F}_2\). From equation \ref{sigma}, the first order derivative of free energy with respect to the strain corresponds to the stress. In the undeformed state, where the stress is zero, the first order term in the expansion does not contribute. The expansion includes two independent second order terms: the squared sum of the diagonal entries and the squared sum of all entries, expressed as:
\begin{equation}
    \mathcal{F} = \mathcal{F}_0 + \frac{1}{2} \lambda \sum_i \epsilon_{ii}^2 + \mu \sum_{i,k} \epsilon_{ik}^2
\end{equation}
where \(\lambda\) and \(\mu\) are known as the Lamé parameters. We can separate the diagonal and off-diagonal parts in the second term using equation \ref{eps_ik}:
\begin{equation}
    \mathcal{F} = \mathcal{F}_0 + \mu \sum_{i,k} \left(\epsilon_{ik} - \frac{1}{3}\delta_{ik} \epsilon_{ll}\right)^2 + \frac{1}{2}K \sum_l \epsilon_{ll}^2
\end{equation}
where \(K\) is called the bulk modulus and is related to the Lamé parameters by \(K = \lambda + \frac{2}{3}\mu\).

\subsubsection{Deformation Due to Change of Temperature}

Suppose the undeformed temperature of a body is \(T_0\). At a different temperature \(T \neq T_0\), the body will deform, leading to the introduction of a first-order term in the expansion of free energy. This term must be linear and scalar, and the only suitable candidate is the trace of the strain tensor. For small deformations, we can assume that the coefficient of this linear term linearly depends on the temperature. Since this first-order term vanishes at \(T = T_0\), we define this coefficient as \(-K\alpha(T-T_0)\), where \(\alpha\) is the thermal expansion coefficient. Consequently, the free energy can be expressed as:
\begin{equation}
\begin{split}
      \mathcal{F} = & \mathcal{F}_0 - K\alpha(T-T_0) \Tr(\epsilon) \\ 
      & + \mu \sum_{i,k} \left(\epsilon_{ik} - \frac{1}{3}\delta_{ik} \epsilon_{ll}\right)^2 + \frac{1}{2}K \sum_l \epsilon_{ll}^2
\end{split}
\end{equation}
To solve equation \ref{weak form}, we describe the stress tensor as a function of the strain tensor, utilizing the differential relationship from equation \ref{sigma}:
\begin{equation}
\begin{split}
     \sigma_{ik} = & -K\alpha(T-T_0)\delta_{ik} \\ 
     & + K \sum_l \delta_{ik}\epsilon_{ll} + 2\mu \left(\epsilon_{ik} - \frac{1}{3}\delta_{ik}\sum_l \epsilon_{ll}\right)
\end{split}
\end{equation}
Notice that the sum symbol over \(i,k\) disappears when taking the differential. When the differential of \(\mathcal{F}\) involves a specific entry of the strain tensor, terms in the sum with different indices are independent of that entry and consequently vanish. For convenience of FEM analysis, we may alternatively express this as:
\begin{equation}
    \mathbf{\sigma} = \lambda \Tr(\mathbf{\epsilon}) \mathbf{I} + 2\mu \mathbf{\epsilon} - \alpha (3\lambda + 2\mu)(T-T_0)\mathbf{I}
\end{equation}
Substituting this expression into equation \ref{weak form}, we derive the final form of the equation to be solved in this study \cite{bleyer2018numericaltours, landau1986theory}.

\section{FEM Strain Simulation Details}\label{FEM Details}

\subsection{Meshing}

\begin{figure}[h!]
    \centering
    \includegraphics[width=1\linewidth]{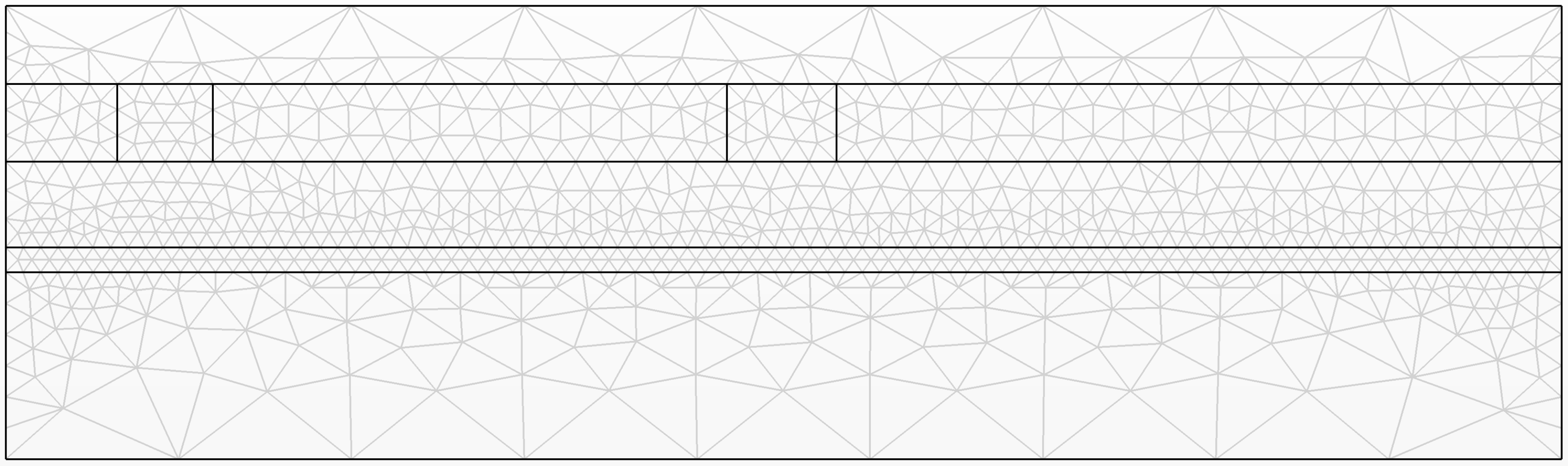}
    \caption{Side view of the simulation mesh, showing how mesh density is controlled using the transfinite feature in Gmsh. Dark lines mark the boundaries between layers: from top to bottom, the \(\text{Al}_2\text{O}_3\) cap layer, the Al layer, the \(\text{Ge}_{0.8}\text{Si}_{0.2}\) substrate, the Ge layer, and another \(\text{Ge}_{0.8}\text{Si}_{0.2}\) substrate. Grey lines show the edges of mesh elements. The bottom surface has 10 nodes of mesh elements per edge, the top surface has 15, and each edge within the germanium layer has 100 nodes. The mesh is deliberately made denser in the germanium layer and around the aluminum gates, where the strain distribution is expected to be more complex and is more relevant to our study.}
    \label{Mesh}
\end{figure}

To solve the thermal elasticity equation on our geometry using the FEM, we partition the domain into mesh grids. Each grid consists of vertices, facets, and cells. The discretized equation is solved at each vertex, with boundary conditions applied to the facets, and the solution is interpolated within the cells based on the element type. For instance, in first-order elements, the solution inside the cell is obtained by linearly interpolating values at the vertices. Increasing the mesh density or using higher-order elements can improve accuracy but at the cost of increased computational time and resources. Since our focus is on the germanium layer where the quantum dots reside, we use a highly refined mesh in this region and reduce mesh density elsewhere to balance accuracy and efficiency. For details of the mesh see Fig \ref{Mesh}. 

\subsection{Weak Coupling}

In a complete treatment, the mechanical and temperature field are coupled: temperature changes distort the material, and material distortion affects the temperature distribution. However, in this study, we assume that the latter effect is negligible. This assumption allows us to solve the mechanical problem using the thermal elasticity equation, Eq. \ref{Elasticity}, without coupling the temperature field to mechanical quantities. The stress tensor in this equation is defined by Eq. \ref{stress}, where the current temperature and the unstrained temperature are treated as constants rather than variables interacting with the mechanical fields. This approach is referred to as weak coupling \cite{bleyer2018numericaltours}. Our device is assembled at room temperature (\(T_0 = \SI{300}{K}\)) and operates at cryogenic temperatures, which we set as \(T = \SI{20}{K}\). 

\subsection{Solver Settings}

We solve the thermal elasticity equation in FEniCS~\cite{Alnaes2015FEniCS}, where the displacement is defined in a first-order Continuous Galerkin (CG) function space. 'First-order' means that the solution within each mesh cell is linearly approximated, while 'CG' indicates that the mesh elements belong to the Lagrange family, ensuring continuity across cell boundaries. After solving for the displacement field, we can obtain strain tensor by taking the gradient of the displacement by Eq. \ref{eps_ik}. Since 3D FEM problems are computationally intensive, selecting an efficient linear solver and preconditioner is essential for improving computational performance.

In this study, we use the GMRES (Generalized Minimal Residual Method) solver with the Hypre-AMG (Algebraic Multigrid) preconditioner. GMRES is an iterative solver particularly suited for solving large, sparse, and non-symmetric systems of linear equations, which are common in finite element analysis of complex PDEs. It minimizes the residual vector's Euclidean norm in a Krylov subspace, expanding this subspace incrementally until the solution converges or a predefined number of iterations is completed \cite{Saad}.

Hypre provides high-performance preconditioners like AMG, which is especially effective for large-scale linear systems. AMG constructs a hierarchy of coarser grids to accelerate convergence. The system is approximately solved on a coarse grid, capturing large-scale dynamics before refining the solution on finer grids \cite{xu2017algebraic}.

\subsection{Parameter Choices}

There are three key parameters that need to be specified to calculate the strain distribution using FEM: the Lamé parameters \(\lambda\) and \(\mu\), and the Thermal Expansion Coefficient (CTE) \(\alpha\). The variation of the Lamé parameters with temperature is negligible, so we use their values at cryogenic temperature \cite{abadillo-uriel2023hole} (see Table \ref{lame_parameters}).
\begin{table}[h!]
    \centering
    \caption{Lamé Parameters (\(\lambda\) and \(\mu\)) for Various Materials}
    \begin{tabular}{lcc}
        \toprule
        Material & \(\lambda\) (GPa) & \(\mu\) (GPa) \\
        \midrule
        Ge               & 49.0  & 68.8 \\
        Ge\(_{0.8}\)Si\(_{0.2}\)  &   52.2   &  71.1  \\
        Al               & 61.4 & 30.9 \\
        Al\(_2\)O\(_3\)  & 63.3  & 68.5 \\
        \bottomrule
    \end{tabular}
    \label{lame_parameters}
\end{table}\\
The thermal expansion coefficient on the other hand, strongly depends on the temperature \cite{slack1975thermal, levinshtein2001properties, schauer1965thermal}. We adopt similar approach as in \cite{frink2024reducing}, which uses the temperature averaged CTE $\bar{\alpha}$ defined as: 
\begin{eqnarray}
    \bar{\alpha} = \left( T_0 - T \right)^{-1} \int_{T}^{T_0} \alpha(T') \, dT'
\end{eqnarray}
The values we use is summarized in Table \ref{tab:thermal_expansion}
\begin{table}[h!]
    \centering
    \caption{Thermal Expansion Coefficients for Various Materials}
    \begin{tabular}{lcc}
        \toprule
        Material & Average CTE $\bar{\alpha}$ ($10^{-6} \text{K}^{-1}$) & Reference \\
        \midrule
        Ge               & 0.76   & \cite{slack1975thermal} \\
        Ge\(_{0.8}\)Si\(_{0.2}\)  & 4.37   & \cite{levinshtein2001properties} \\
        Al               & 14.16  & \cite{frink2024reducing} \\
        Al\(_2\)O\(_3\)  & 3.30   & \cite{schauer1965thermal} \\
        \bottomrule
    \end{tabular}
    \label{tab:thermal_expansion}
\end{table}
According to Eq. \ref{g-correction}, to calculate the g-tensor correction from strain, we only need the deformation potentials of germanium since quantum dots only occur in the germanium layer. Their values are summarized in Table \ref{deformation_potentials} \cite{abadillo-uriel2023hole}.
\begin{table}[h!]
    \centering
    \caption{Hydrostatic, Uniaxial, and Shear Deformation Potentials for Germanium}
    \begin{tabular}{lc}
        \toprule
        Parameter & Value (eV) \\
        \midrule
        Hydrostatic Deformation Potential (\(a_v\))  &  2.00\\
        Uniaxial Deformation Potential (\(b_v\))     &  -2.16\\
        Shear Deformation Potential (\(d_v\))        &  -6.06\\
        \bottomrule
    \end{tabular}
    \label{deformation_potentials}
\end{table}

\subsection{Boundary Condition}

In Section \ref{Device Structure}, we explained that we limited the simulation to a portion of the device to focus on the area containing the quantum dots. This selection introduced subtle considerations for the choice of boundary conditions. Ideally, the boundary conditions should accurately reflect the displacement vector distribution along the boundary of the simulation geometry. We tested two boundary conditions: the first fixed all outer surfaces, while the second fixed only the top and bottom surfaces, leaving the sides free. We selected the first option for our final results. To validate this choice, we created a geometry similar to that in \cite{abadillo-uriel2023hole} and found that the first condition reproduced the same pattern and order of magnitude as reported in that study. In contrast, the second condition produced a significantly different pattern, though it remained within the same order of magnitude. 

\newpage

\end{document}